\documentclass[useAMS,usenatbib]{mn2e}
\pdfoutput=1
\usepackage{aas_macros}
\usepackage{epsfig}
\usepackage{amsmath}
\usepackage{amssymb}
\usepackage{leftidx}
\usepackage{natbib}
\usepackage[rgb]{xcolor}
\definecolor{MyGreen}{rgb}{0.0,0.6,0.3}
\definecolor{MyPurple}{rgb}{0.6,0,0.3}
\usepackage{fancyref}
\usepackage{hyperref}
\hypersetup{colorlinks=true,citecolor=MyGreen,linkcolor=MyPurple,urlcolor=blue}

\def\beq{\begin{equation}}
\def\eeq{\end{equation}}
\def\ba{\begin{eqnarray}}
\def\ea{\end{eqnarray}}
\def\bal{\begin{align}}
\def\eal{\end{align}}
\def\bxi{{\mbox{\boldmath $\xi$}}}

\begin{document}

\title[Pre-Supernova Outbursts] {Pre-Supernova Outbursts via Wave Heating in Massive Stars I: Red Supergiants}

\author[Fuller et al.]{
Jim Fuller$^{1,2}$\thanks{Email: jfuller@caltech.edu}\\
\\$^1$TAPIR, Walter Burke Institute for Theoretical Physics, Mailcode 350-17, Caltech, Pasadena, CA 91125, USA
\\$^2$Kavli Institute for Theoretical Physics, Kohn Hall, University of California, Santa Barbara, CA 93106, USA}

\label{firstpage}
\maketitle

\begin{abstract}

Early observations of supernovae (SNe) indicate that enhanced mass loss and pre-SN outbursts may occur in progenitors of many types of SNe. We investigate the role of energy transport via waves driven by vigorous convection during late-stage nuclear burning of otherwise typical $15 \, M_\odot$ red supergiant SNe progenitors. Using MESA stellar evolution models including 1D hydrodynamics, we find that waves carry $\sim \! 10^7 \, L_\odot$ of power from the core to the envelope during core neon/oxygen burning in the final years before core collapse. The waves damp via shocks and radiative diffusion at the base of the hydrogen envelope, which heats up fast enough to launch a pressure wave into the overlying envelope that steepens into a weak shock near the stellar surface, causing a mild stellar outburst and ejecting a small ($\lesssim 1 \, M_\odot$) amount of mass at low speed ($\lesssim \! 50 \, {\rm km}/{\rm s}$) roughly one year before the SN. The wave heating inflates the stellar envelope but does not completely unbind it, producing a non-hydrostatic pre-SN envelope density structure different from prior expectations. In our models, wave heating is unlikely to lead to luminous type IIn SNe, but it may contribute to flash-ionized SNe and some of the diversity seen in II-P/II-L SNe.

\end{abstract}


\section{Introduction}

The connection between the diverse population of core-collapse supernovae (SNe) and their massive star progenitors is of paramount importance for the fields of both SNe and stellar evolution. Over the past decade, substantial evidence has emerged for enhanced pre-SN mass loss and outbursts in the progenitors of several types of SNe. The inferred mass loss rates are typically orders of magnitude larger than those measured in local group massive stars, and the mass loss appears to systematically occur in the last centuries, years, or weeks of the stars' lives. This deepening mystery cannot be explained by standard stellar evolution/wind theories, and its solution lies at the heart of the SNe-massive star connection.

Type IIn SNe provide the most obvious evidence for pre-supernova mass loss, and it is well known that these SNe are powered by interaction between the supernova ejecta and dense circumstellar material (CSM). However, type IIn SNe are very heterogeneous (\citealt{smith:16b} classifies them into ten subtypes), as some appear to require interaction with $\sim \! \! 10 \, M_\odot$ of CSM ejected in the final years of their progenitor's life, while others require mass loss rates of only $\sim \! 10^{-4} \, M_\odot/{\rm yr}$ but lasting for centuries before the explosion \citep{smith:16a}. These mass loss rates are much larger than predicted by standard  mass-loss prescriptions. In some cases, pre-SN outbursts resulting in mass ejection have been observed directly, famous examples being SN 2009ip (which did not explode until 2012, \citealt{mauerhan:13,margutti:14,graham:14,smith:14}), 2010mc \citep{ofek:13b}, LSQ13zm \citep{tartaglia:16}, and SN 2015bh \citep{elias-rosa:16,thone:16,ofek:16}, which show resemblance with luminous blue variable (LBV) star outburts. Pre-SN outbursts now appear to be common for type IIn SNe \citep{ofek:14}.


Enhanced pre-SN mass loss has also been inferred from observations of other types of SNe. Type Ibn SNe (e.g., SN 2006jc which had a pre-SN outburst, \citealt{pastorello:07}; and SN 2015U, \citealt{shivvers:16}) show interaction with He-rich material ejected soon before core-collapse.
SN 2014C was a type Ib SNe that transitioned into a type IIn SNe after the ejecta collided with a dense shell of H-rich CSM ejected by its progenitor in its final $\sim$decades of life \citep{milisavljevic:15,margutti:16}. Early spectra of type IIb SN 2013cu reveal emission lines from a flash-ionized wind \citep{gal-yam:14} with inferred mass loss rates over $10^{-3} \, M_\odot/{\rm yr}$ \citep{groh:14}. Many bright type II-P/II-L SNe also show flash-ionized emission lines in early time spectra indicative of a thick stellar wind \citep{khazov:16}, while even relatively normal II-P SNe sometimes exhibit peaks in their early light curves that may be produced by shock cooling of an extremely dense stellar wind \citep{moriya:11,morozova:16}. Recently, \cite{yaron:17} found that the otherwise normal type II-P SN2013fs showed emission lines only within the first several hours after explosion, indicating that modest mass ejection of $\sim \! 10^{-3} \, M_\odot$ in the final $\sim$year of the progenitor's life is common for type II-P SNe.

\begin{figure}
\begin{center}
\includegraphics[scale=0.32]{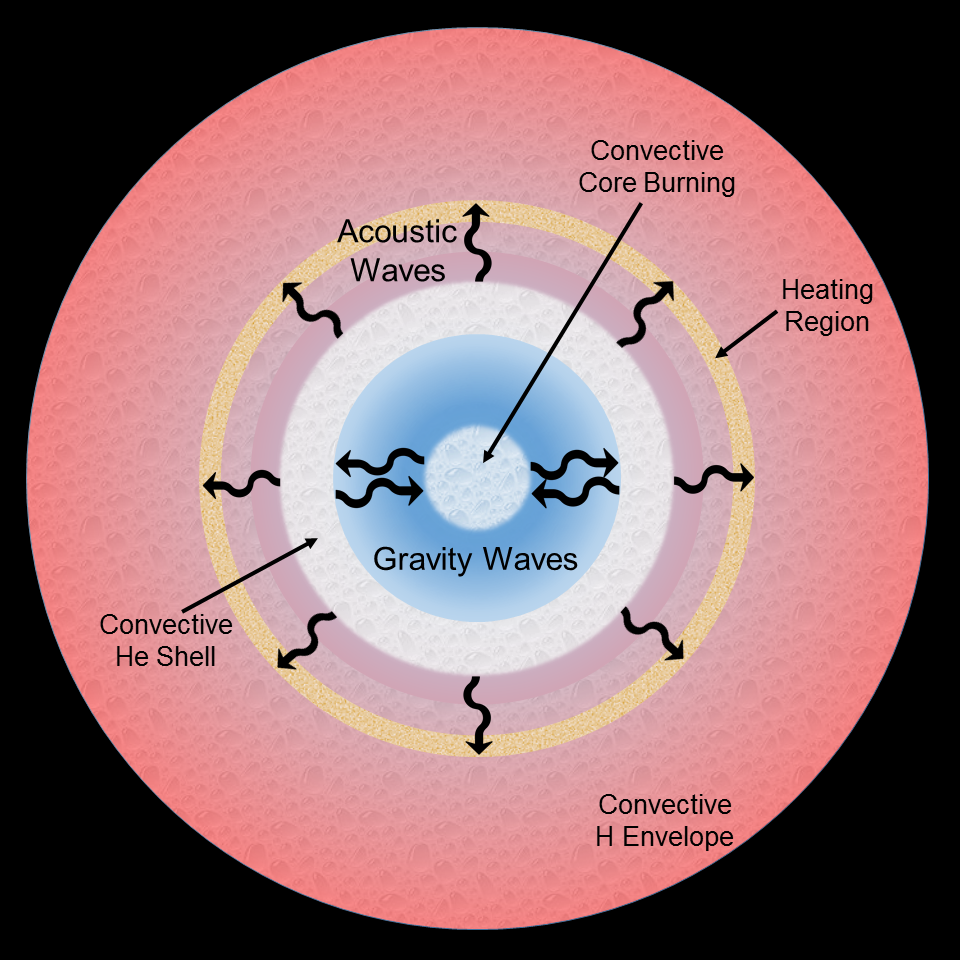}
\end{center} 
\caption{ \label{fig:Cartoon} 
Cartoon (not to scale) of wave heating in a red supergiant. Gravity waves are excited by vigorous core convection and propagate through the outer core. After tunneling through the evanescent region created by the convective He-burning shell, they propagate into the H envelope as acoustic waves. The acoustic waves damp near the base of the envelope and heat a thin shell.}
\end{figure}

One of the most promising explanations for pre-SN outbursts and mass loss was proposed by \cite{quataert:12}, who investigated the impacts of convectively driven hydrodynamic waves during late-phase nuclear burning. Convectively driven waves are a generic consequence of convection that are routinely observed in hydrodynamic simulations. \cite{quataert:12} showed that the vigorous convection of late burning stages (especially Ne/O burning) can generate waves carrying in excess of $10^7 \, L_\odot$ of power to the outer layers of the stars, potentially depositing more than $10^{47} \, {\rm erg}$ in the envelope of the star over its last months/years of life. \autoref{fig:Cartoon} provides a cartoon picture of the wave heating process. \cite{shiode:14} then showed that the wave heating is generally more intense but shorter-lived in more massive stars, and could occur in a variety of SN progenitor types. More recently, \cite{quataert:16} have examined the effect of super-Eddington heat deposition (e.g., due to wave energy) near the surface of a star, showing that the heat can drive a dense wind with a very large mass-loss rate.

In this paper, we examine wave heating effects in otherwise ``typical" $M_{\rm ZAMS} \! = \! 15 \, M_\odot$ red supergiants (RSGs) that may give rise to type II-P, II-L, or IIn supernovae depending on the impact of wave heating. We quantify how wave heating alters the stellar structure, luminosity, and mass-loss rate using MESA simulations \citep{paxton:11,paxton:13,paxton:15} including the effects of wave heating due to convectively driven waves. After carbon shell burning, we use the 1D hydrodynamic capabilities of MESA to account for the pressure waves, shocks, and hydrodynamic/super-Eddington mass loss that can result from intense wave heating.


\section{Wave Energy Transport}
\label{imp}

\subsection{Wave Generation}

\begin{figure}
\begin{center}
\includegraphics[scale=0.36]{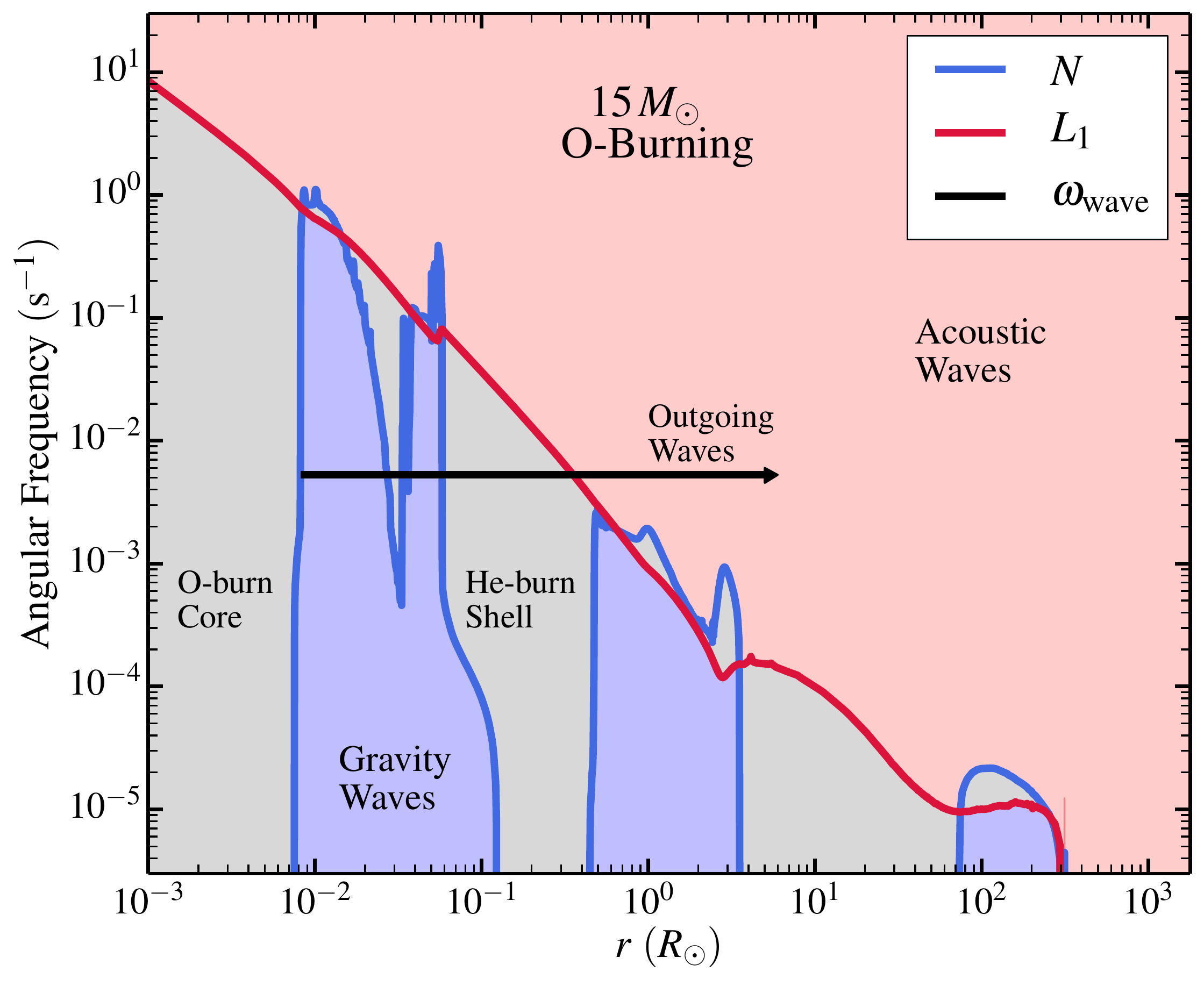}
\end{center} 
\caption{ \label{fig:15MsunProp} 
Propagation diagram for our model during core oxygen burning, showing the Brunt-V\"{a}is\"{a}l\"{a} frequency $N$ and the $\ell=1$ Lamb frequency $L_1$. Vigorous convection in the core excites waves of frequency $\omega_{\rm wave} \! \sim 5 \! \times \! 10^{-3} \, {\rm rad}/{\rm s}$ that propagate through the core as gravity waves. The waves must tunnel through one or two evanescent zones before penetrating into the stellar envelope as acoustic waves, where their energy is dissipated into heat.}
\end{figure}

Gravity waves are low frequency waves that can propagate in radiative regions of stars where their angular frequency $\omega$ is smaller than the Brunt-V\"{a}is\"{a}l\"{a} frequency $N$ (see \autoref{fig:15MsunProp}).
They are excited at the interface between convective and radiative zones, carrying energy and angular momentum into the radiative zone which is sourced from the kinetic energy of turbulent convection. The energy carried by gravity waves is a small fraction of the convective luminosity, scaling roughly as \citep{goldreich:90}
\beq
\label{Lwave}
L_{\rm wave} \sim \mathcal{M}_{\rm con} L_{\rm con} \, ,
\eeq
where $L_{\rm con}$ is the luminosity carried by convection and $\mathcal{M}_{\rm con}$ is a typical turbulent convective Mach number. In most phases of stellar evolution, $\mathcal{M}_{\rm con} \! \lesssim \! 10^{-3}$ within interior convection zones, and the energy carried by gravity waves is negligible. Equation \ref{Lwave} has been approximately verified by multidimensional simulations (\citealt{rogers:13,rogers:15,alvan:14,alvan:15}).



\begin{figure}
\begin{center}
\includegraphics[scale=0.32]{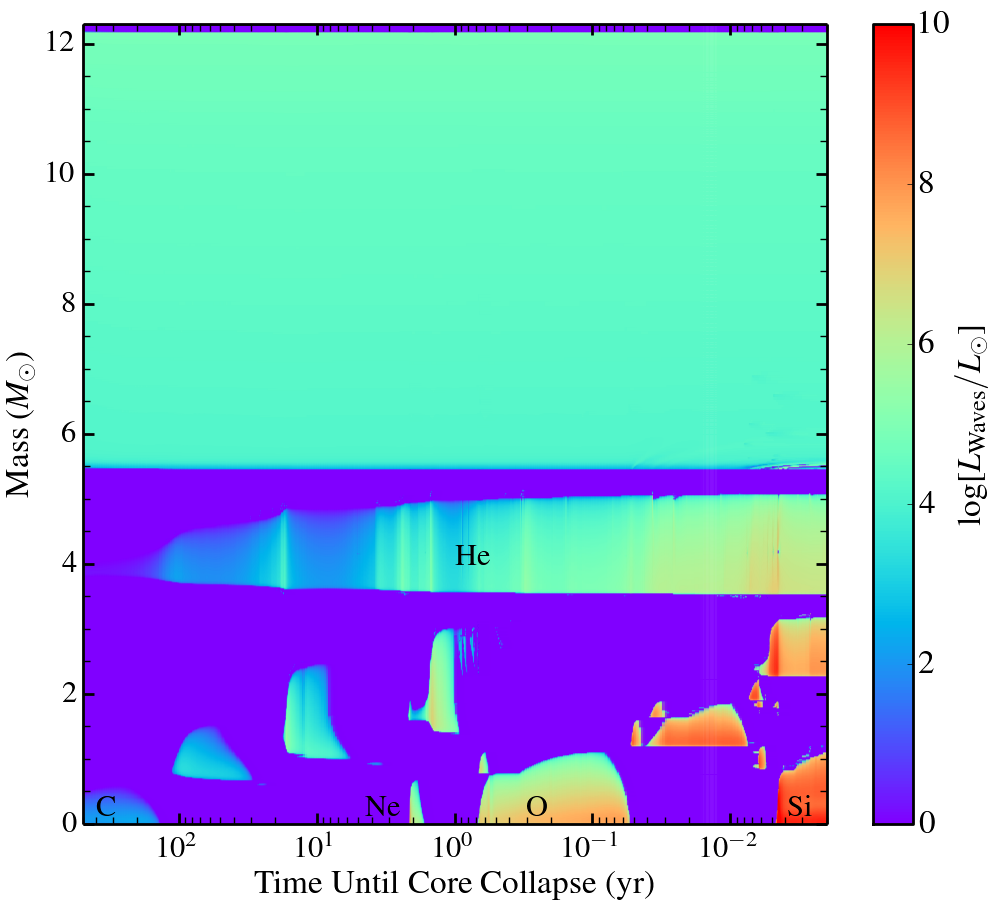}
\end{center} 
\caption{ \label{fig:15MsunKipp} 
Kippenhahn diagram of our $M_{\rm ZAMS} = 15 \, M_\odot$ model from carbon burning through silicon burning. Shading indicates the wave energy luminosity $L_{\rm wave} = \mathcal{M}_{\rm con} L_{\rm con}$ each convective zone is capable of generating, and zones are labeled by the element they burn. Purple regions are stably stratified regions where convectively excited gravity waves may propagate.}
\end{figure}


\autoref{fig:15MsunKipp} shows the quantity $L_{\rm wave}$ within the interior of a $M_{\rm ZAMS} = 15 \, M_\odot$ stellar model from core carbon burning onward. Details and parameters of our MESA models can be found in \autoref{models}. Before carbon shell burning, $L_{\rm wave}$ is much less than the surface luminosity of $L \simeq 10^5 \, L_\odot$, and wave energy transport is negligible. However, after carbon burning, neutrino cooling becomes very efficient within the core, which falls out of thermal equilibrium with the envelope. To maintain thermal pressure support, burning luminosities increase and become orders of magnitude larger than the surface luminosity. Convective mach numbers also increase, and consequently $L_{\rm wave}$ during late burning phases can greatly exceed the surface luminosity, allowing wave energy redistribution to produce dramatic effects.

To estimate wave luminosities in our 1D models, we proceed as follows. First, we calculate $L_{\rm wave}$ at each radial coordinate as shown in \autoref{fig:15MsunKipp}. Next, we calculate a characteristic convective turnover frequency at each radial coordinate via 
\beq
\label{omegaconv}
\omega_{\rm con} = 2 \pi \frac{v_{\rm con}}{2 \alpha_{\rm MLT} H} \, ,
\eeq
where
\beq
\label{vconv}
v_{\rm con}=\big[L_{\rm con}/(4 \pi \rho r^2)\big]^{1/3} \, ,
\eeq
is the RMS convective luminosity according to mixing length theory (MLT), $\alpha_{\rm MLT}$ is the mixing length parameter, and $H$ is a pressure scale height. The turbulent mach number is $\mathcal{M}_{\rm con} = v_{\rm con}/c_s$, where $c_s$ is the adiabatic sound speed. Remarkably, these estimates of convective velocities and turnover frequencies typically match those seen in 3D simulations of a variety of burning phases (e.g., \citealt{meakina:07,alvan:14,couch:15,lecoanet:16,jones:17}) to within a factor of two.

In reality, a spectrum of waves with different angular frequencies $\omega$ and angular wavenumbers $k_\perp = \sqrt{l(l+1)}/r$ are excited by each convective zone, where $l$ is the spherical harmonic index of the wave. Rather than model the wave spectrum, we find the maximum value of $\omega_{\rm con}$ (usually located a fraction of a scale height below the zone's outer radius), and assume that all the wave power is put into waves at this frequency
\beq
\label{omegawave}
\omega_{\rm wave} = \omega_{\rm con,max} \, ,
\eeq
and angular wave numbers $l=1$. Simulations show that realistic wave spectra are peaked around $\omega = \omega_{\rm wave}$ and $l=1$, even for fairly thin shell convection like that in the Sun (see \citealt{alvan:14}), at least for waves not immediately damped, so these approximations are reasonable. Waves at lower frequencies are typically much more strongly damped, while waves at higher frequencies contain much less power. Waves at higher values of $l$ contain comparable or less power and are more strongly damped, so we ignore their contribution. At each time step in our simulations, we find the radial location of $\omega_{\rm max}$ within the core (usually located within the innermost convective burning zone), and then compute $v_{\rm con}$, $\omega_{\rm wave}$, and $L_{\rm wave}$ at that point using equations \ref{Lwave}, \ref{omegaconv}, and \ref{vconv}.

\subsection{Wave Propagation and Dissipation}

The next step is to calculate how waves of frequency $\omega_{\rm wave}$ and $l=1$ will propagate and dissipate within the star. Typical waves at $\omega=\omega_{\rm wave}$ during late burning phases are gravity waves in the core of the star, but in the envelope they are acoustic waves (see \autoref{fig:15MsunProp}). In order to propagate into the envelope, the waves must tunnel through one or more intervening evanescent zones, the largest of which is often created by the convective helium burning shell. Apart from wave evanescence, we ignore wave interactions with convection in these regions because their convective energy fluxes and turnover frequencies are generally much smaller than the core convection that launches the waves, although some interaction may take place. Before tunneling out of the core, the waves may reflect multiple times and can be damped by neutrino emission or by breaking near the center of the star, dissipating some of their energy within the core. In \autoref{waveprop}, we provide details of how to calculate these effects in order to determine the fraction of wave energy $f_{\rm esc}$ which is able to escape from the core and propagate into the envelope as acoustic waves.

The wave energy that heats the envelope is then
\beq
\label{waveheat}
L_{\rm heat} = \eta f_{\rm esc} L_{\rm wave} \, .
\eeq
Here, $\eta$ is an efficiency parameter (with nominal value $\eta =1$ unless stated otherwise) we will adjust to explore the dependence of our results on the somewhat uncertain wave flux. We find typical values of $f_{\rm esc} \! \sim \! 0.5$ during core neon/oxygen burning, and $f_{\rm esc} \! \sim \! 0.1$ during shell burning phases because more wave energy is lost by tunneling into the core. We do not compute the effect of wave heating within the core because its binding energy is much larger than integrated wave heating rates, and because neutrinos can efficiently remove much of this thermal energy.

\begin{figure}
\begin{center}
\includegraphics[scale=0.36]{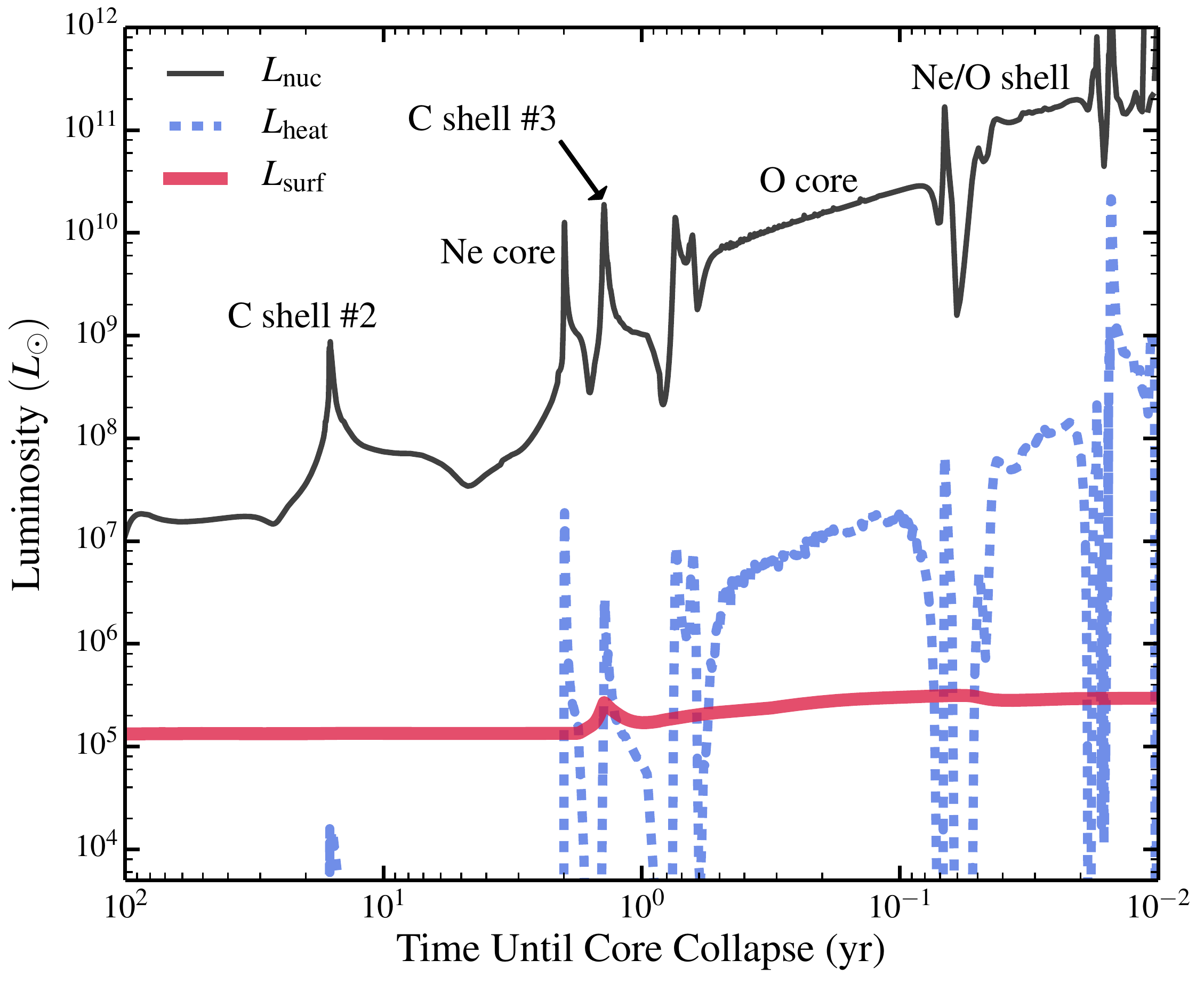}
\end{center} 
\caption{ \label{fig:Luminosity15Msun} 
Luminosity of our $M_{\rm ZAMS}=15 \, M_\odot$ stellar model in its final century before core-collapse. The red line shows the observable surface luminosity, while the black line is the nuclear energy generation rate. A small fraction of this energy is converted into waves which propagate out of the core. The value of $L_{\rm heat}$ is the wave heating rate at the base of the hydrogen envelope.}
\end{figure}

\autoref{fig:Luminosity15Msun} shows the nuclear energy generation rate $L_{\rm nuc}$ (not including energy carried away by neutrinos) of our stellar model as a function of time, along with the envelope wave heating rate $L_{\rm heat}$ and the surface luminosity $L_{\rm surf}$.  Important burning phases are labeled. Although the fraction of nuclear energy converted into waves that escape the core is generally very small ($<10^{-3}$), the value of $L_{\rm heat}$ can greatly exceed $L_{\rm surf}$. In our models, $L_{\rm surf}$ remains smaller than $L_{\rm heat}$ during later burning phases because most of the wave heat remains trapped under the H envelope and is not radiated by the photosphere, which we discuss more in \autoref{effects}.

\begin{figure}
\begin{center}
\includegraphics[scale=0.34]{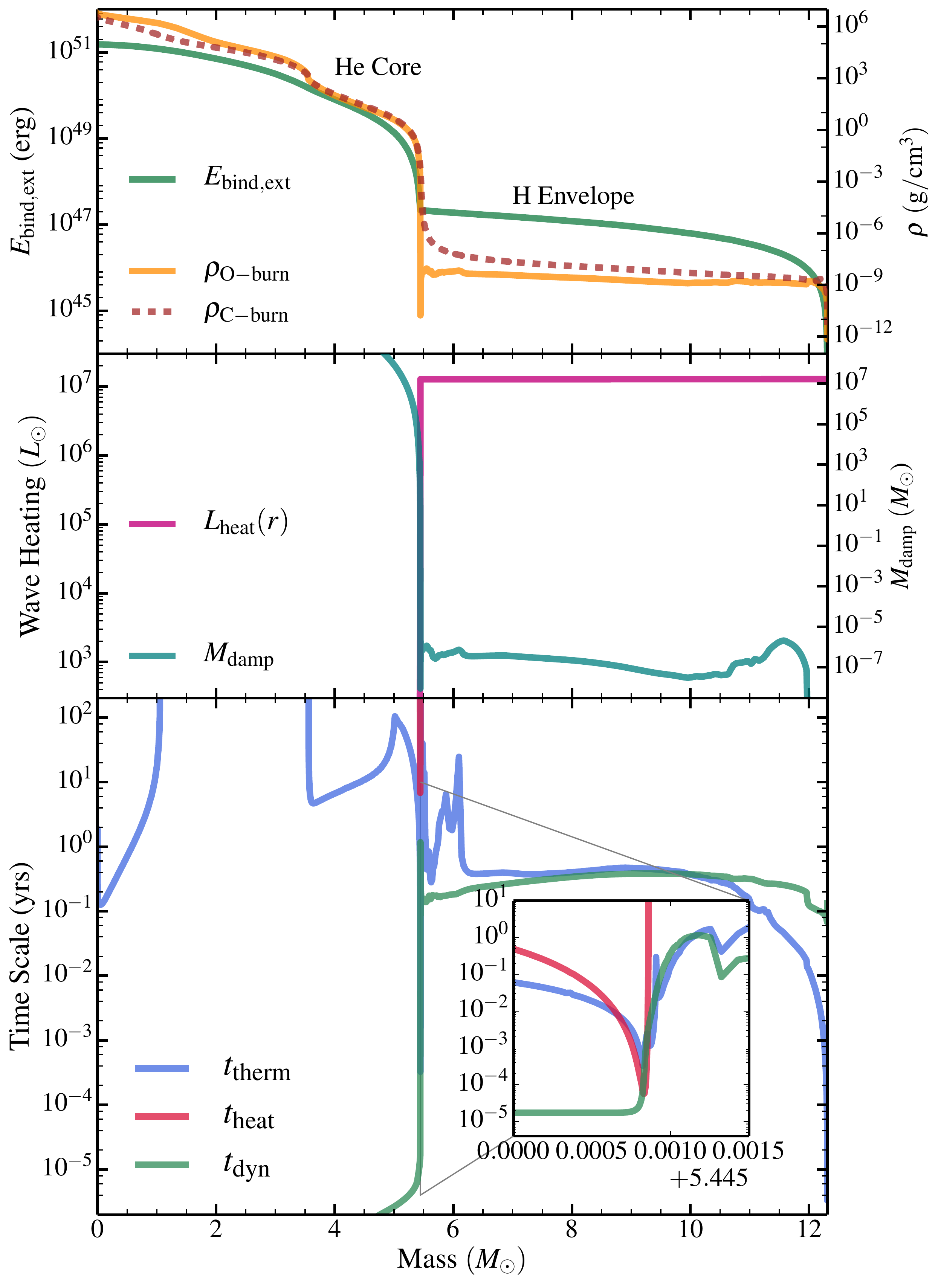}
\end{center} 
\caption{ \label{fig:WaveDamp15Msun} 
{\bf Top:} Binding energy integrated inward from the surface of our $M_{\rm ZAMS} = 15 \, M_\odot$ model just after carbon burning, as function of mass coordinate. The right axis shows the corresponding density profile just after carbon burning, and during oxygen burning. {\bf Middle:} Wave heating rate $L_{\rm heat}(r)$, integrated from the center of the star to the local mass coordinate, during oxygen burning. Essentially all of the wave heat is deposited at the base of the hydrogen envelope at mass coordinate $m \simeq 5.446 \, M_\odot$. The right axis shows the damping mass $M_{\rm damp}$ through which the waves must propagate to be attenuated (equation \ref{Mdamp}). $M_{\rm damp}$ plummets just outside the core, causing the waves to damp at that location. {\bf Bottom:} Dynamical, thermal, and wave heating timescales as defined in \autoref{effects}. The long thermal timescale above the heating region prevents most wave heat from diffusing outward. Wave heating causes these timescales to be very short and comparable to one another in the heating region (inset).}
\end{figure}

After determining $L_{\rm heat}$, we must determine where within the envelope the wave energy will damp into thermal energy. This calculation is detailed in \autoref{wavedamp}, where we calculate wave damping via thermal diffusion and describe how we add wave heat into our stellar model. The most important feature of diffusive wave damping is that it is strongly dependent on density and sound speed, with a characteristic damping mass $M_{\rm damp} \! \propto \! \rho^{3}$ (equation \ref{Mdamp}). In RSGs, the density falls by a factor of $\sim \! \! 10^6$ from the helium core to the base of the hydrogen envelope (see \autoref{fig:WaveDamp15Msun}). Hence, acoustic waves at frequencies of interest are essentially undamped in the helium core but quickly damp as they propagate into the hydrogen envelope, and they always thermalize their energy in a narrow shell of mass at the base of the hydrogen envelope.

\begin{figure}
\begin{center}
\includegraphics[scale=0.36]{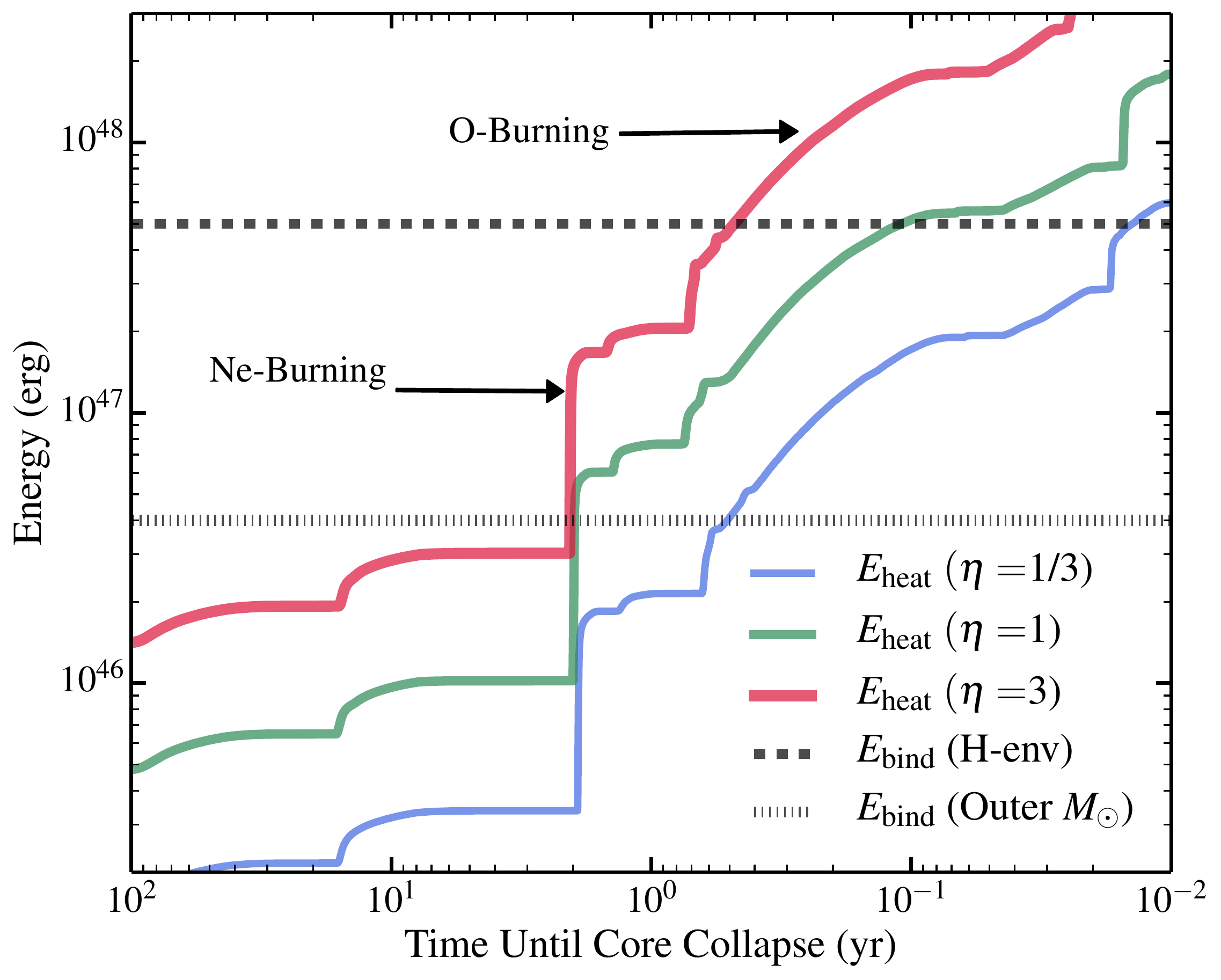}
\end{center} 
\caption{ \label{fig:Edep15MsunComp} 
Integrated wave energy deposited outside of the core (starting from core carbon burning) as a function of time until core collapse, for three different heating efficiencies $\eta$. The dashed black line shows the total binding energy of the hydrogen envelope (in a model not including wave heating). The dotted black line is the binding energy of the outer solar mass of the envelope (see \autoref{fig:WaveDamp15Msun}).}
\end{figure}

In the late stages of preparing this manuscript, \cite{ro:17} demonstrated that acoustic waves will generally steepen into shocks before damping diffusively, causing them to thermalize their energy deeper in the star. Using their equation 6 and calculating wave amplitudes from the value of $L_{\rm heat}$, we find shock formation in our models occurs at somewhat larger (by a factor of a few) density than radiative diffusion, but at very similar mass coordinates and overlying binding energies. The reason is that the density cliff at the edge of the He core promotes both shock formation and diffusion. We therefore suspect that wave energy thermalization via shock formation will only marginally affect our results, but we plan to account for it in future work.

Our wave heating calculations during shell Ne/O burning and core Si burning are less reliable due to an inadequate nuclear network in our models, and increasing wave non-linearity. These burning phases occur less than an envelope dynamical time before core collapse, giving waves little time to alter envelope structure. For these reasons, we do not closely examine these phases in this work, but large wave luminosities during these phases may affect some progenitors.


\section{Effects on Pre-Supernova Evolution}
\label{effects}

In our models, wave heating is most important during late C-shell burning, core Ne burning, and core O burning. To quantify the effects of wave heating on the pre-SN state of the stellar progenitor, we construct MESA models and evolve them from the main sequence to core-collapse. At each time step, we add wave heat $L_{\rm heat}$ as described in \autoref{imp} and \autoref{waveprop}. Just before C burning, we utilize the 1D hydrodynamic capabilities of MESA (see \autoref{models}) which is essential for capturing the non-hydrostatic dynamics that result from wave heating.

\begin{figure}
\begin{center}
\includegraphics[scale=0.37]{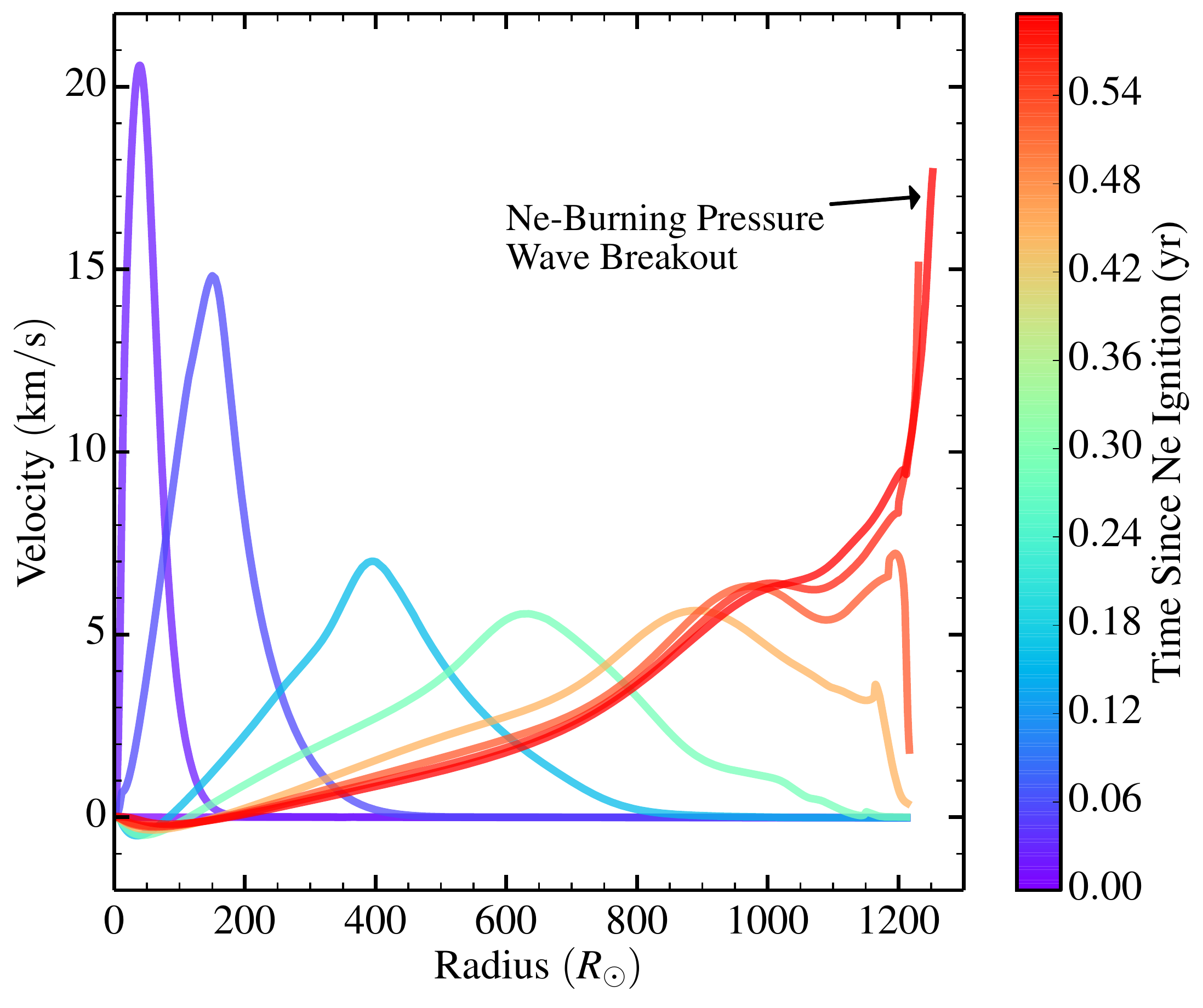}
\end{center} 
\caption{ \label{fig:15MsunVelocity} 
Internal radial velocity profiles of our model at several times measured from the start of core Ne burning. The moving velocity peak arises from the pressure wave that propagates toward the stellar surface, steepening into a weak shock near the photosphere. This weak shock breakout creates the mild outburst shown in Figures \ref{fig:HR15MsunComp} and \ref{fig:TeffR15Msun}. Surface velocities are smaller than the escape speed ($v_{\rm esc} \! \sim \! 45 \, {\rm km}/{\rm s}$), so the surface expands but remains bound.}
\end{figure}



Relative timescales are important for understanding wave heating effects. We define a local wave heating timescale 
\beq
\label{theat}
t_{\rm heat} = \frac{c_s^2}{\epsilon_{\rm heat}} \,
\eeq
where $\epsilon_{\rm heat}$ is the wave heat deposited per unit mass and time. This can be compared with a thermal cooling timescale
\beq
\label{ttherm}
t_{\rm therm} = \frac{4 \pi \rho r^2 H c_s^2}{L} \,
\eeq
where $H$ is the pressure scale height and $L$ is the local luminosity. We also define a local dynamical time scale 
\beq
\label{tdyn}
t_{\rm dyn} = \frac{H}{c_s} \, .
\eeq
Finally, all of these should be considered in relation to the time until core-collapse, $t_{\rm col}$.

\begin{figure*}
\begin{center}
\includegraphics[scale=0.41]{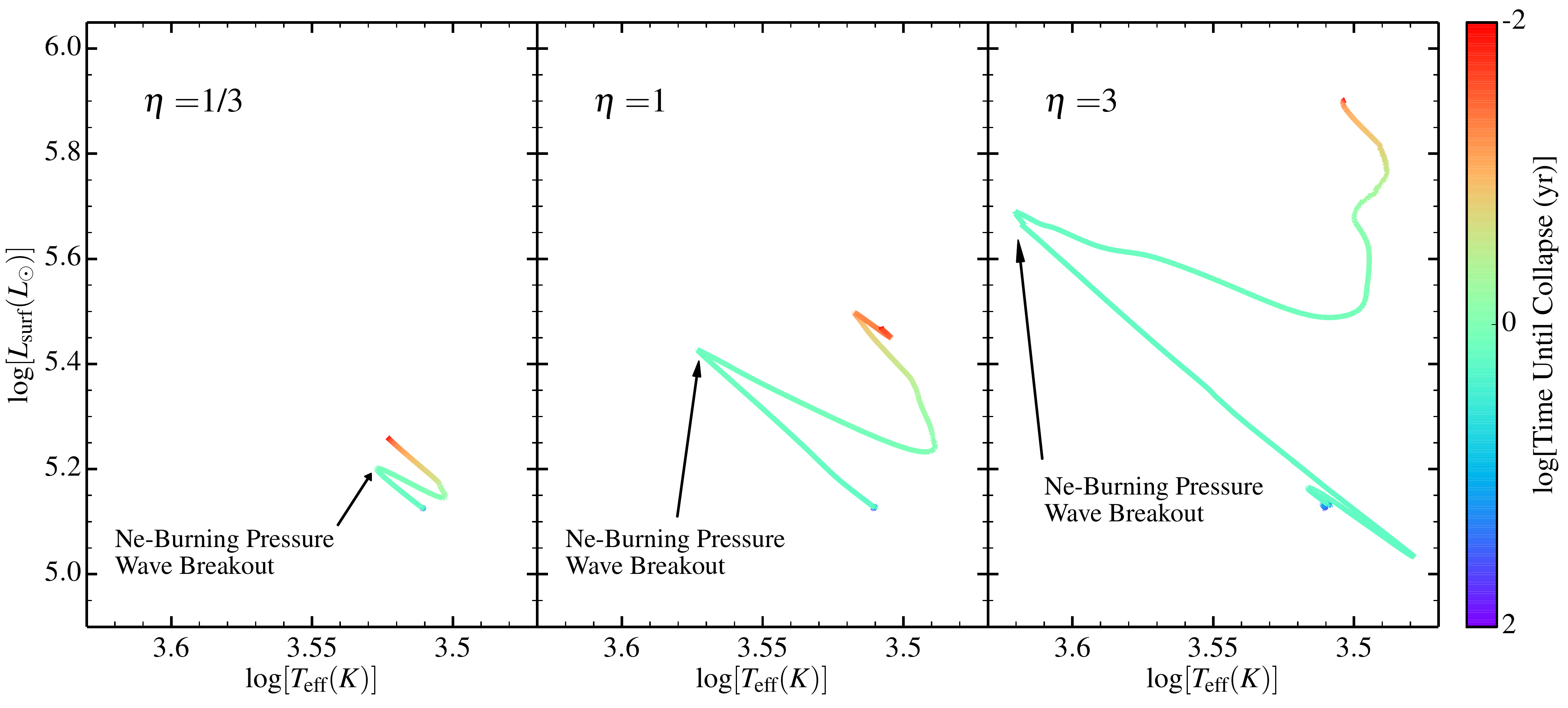}
\end{center} 
\caption{ \label{fig:HR15MsunComp} 
HR diagrams of our models during the century before core-collapse, for different heating efficiencies $\eta$. Stronger wave heating induces stronger surface shock breakouts, creating more dramatic temperature/luminosity increases. }
\end{figure*}

The first key insight is that wave energy is deposited at the base of the hydrogen envelope, above which $t_{\rm therm}$ is comparable to (but generally larger than) $t_{\rm col}$ (see \autoref{fig:WaveDamp15Msun}). Consequently, wave heat cannot be thermally transported to the stellar surface before core-collapse, and the surface luminosity $L_{\rm surf}$ is only modestly affected (\autoref{fig:Luminosity15Msun}). We therefore do not expect very luminous ($L \gtrsim 10^6 \, L_\odot$) pre-SN outbursts to be driven by wave heating in RSGs.

The second key insight is that wave heating timescales can be very short. In the slow heating regime with $t_{\rm heat} \gtrsim t_{\rm therm} \gtrsim t_{\rm dyn}$, wave heat can be thermally transported outward without affecting the local pressure. In the moderate heating regime with $t_{\rm therm} \gtrsim t_{\rm heat} \gtrsim t_{\rm dyn}$, wave heat cannot be thermally transported outward, but the star can expand nearly hydrostatically to accomodate the increase in pressure (see discussion in \citealt{mcley:14}). However, we find wave heating can be so intense that it lies in the dynamical regime $t_{\rm heat} \lesssim t_{\rm therm},t_{\rm dyn}$. In this case, wave heat and pressure build within the wave damping region, exciting a pressure wave which propagates outward at the sound speed (\autoref{fig:15MsunVelocity}). This pressure wave crosses the stellar envelope on a global dynamical timescale
\beq
\
t_{\rm dyn,glob} \sim \sqrt{ \frac{R^3}{GM} } \simeq 0.5 \, {\rm yr}
\eeq
for our stellar model.

\begin{figure}
\begin{center}
\includegraphics[scale=0.33]{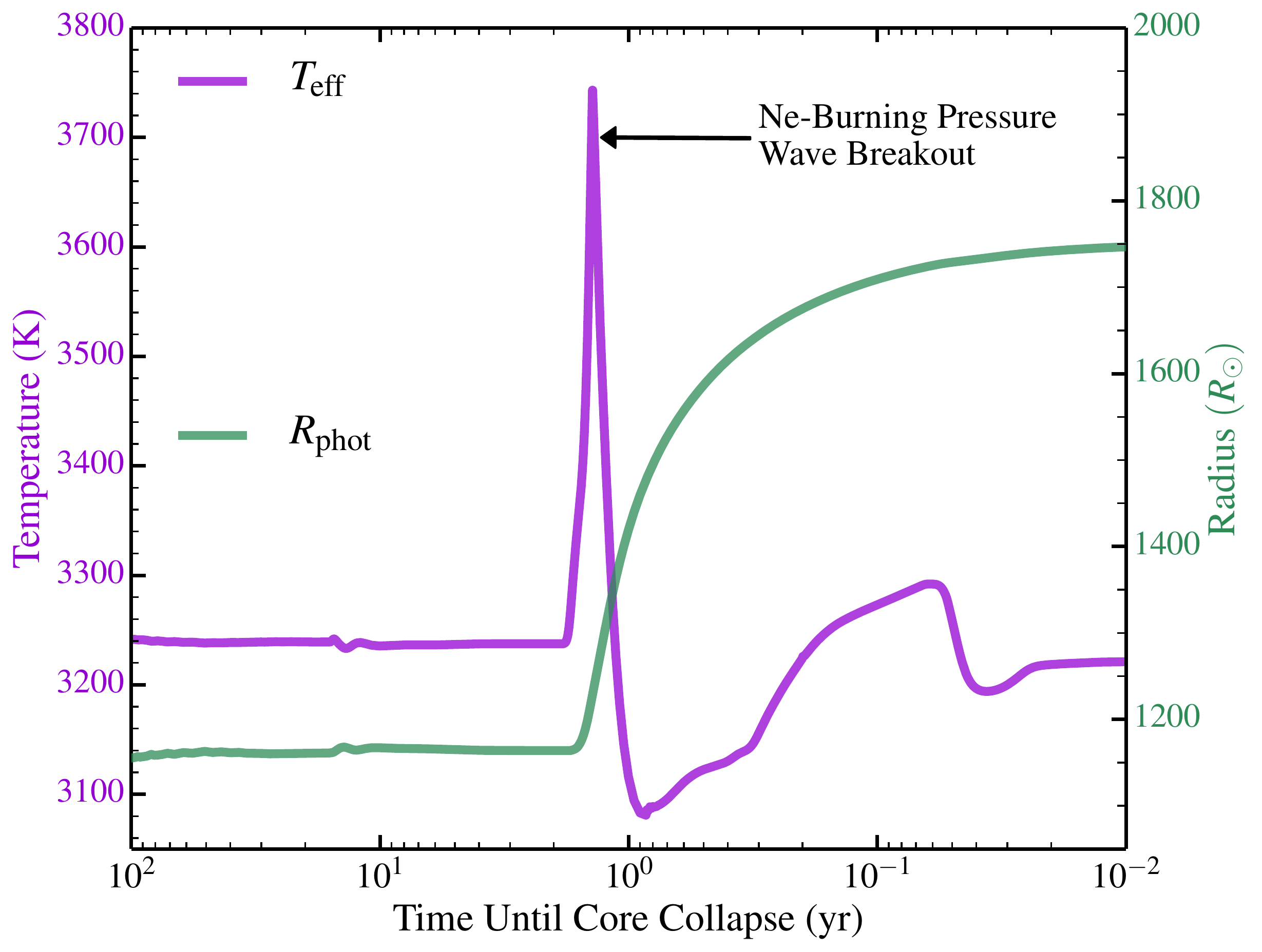}
\end{center} 
\caption{ \label{fig:TeffR15Msun} 
Evolution of the surface temperature and photospheric radius of our stellar model in its final century. The peak in temperature is produced by the Ne-burning wave heating shock breakout, followed by subsequent envelope expansion and cooling. The second, smaller peak is caused by wave heating during late C-shell burning.}
\end{figure}

In our models, the most important envelope pressure wave arises from wave heating during core Ne burning and a third C-shell burning phase (later waves do not reach the surface before core-collapse). As these pressure waves approach the surface where the density and the sound speed drop, they steepen into a weak shock ($\mathcal{M} \lesssim 3$). When the shock wave breaks out of the surface, it produces a sudden spike in surface temperature and luminosity (see Figures \ref{fig:HR15MsunComp} and {\ref{fig:TeffR15Msun}), akin to SN shock breakout \citep{dessart:13} but with much smaller energy, $E \! \sim \! 10^{47} \, {\rm erg}$. This shock breakout is similar to that expected from failed SNe in RSGs \cite{lovegrove:13,piro:13}, but even less energetic and luminous, and preceding core-collapse by months or years. Unlike SNe or failed SNe, the shock in our models is not strong enough to unbind the entire RSG envelope, but it can still drive a small outflow ($M_{\rm out} \! \lesssim \! 1 \, M_\odot$, see \autoref{fig:StructureComp15Msun}) with speeds comparable to  the escape speed $v_{\rm esc}$. After shock breakout, the envelope expands and cools, but is not able to settle back to its quiescent state before core-collapse, or before a subsequent pressure wave is launched by a later burning phase. 

\autoref{fig:HR15MsunComp} shows the evolution of our model in the HR diagram during its last century. The pressure wave breakout creates a jump in surface temperature and luminosity followed by envelope expansion and cooling. The rebrightening just before core-collapse occurs as a second pressure wave (driven by wave heating during C-shell burning) approaches the photosphere. \autoref{fig:TeffR15Msun} shows the corresponding evolution in surface temperature and photospheric radius.

Core O-burning produces a markedly different result from Ne-burning because the wave heating is both stronger and lasts longer, depositing nearly an order of magnitude more energy into the envelope (\autoref{fig:Edep15MsunComp}). In our models, the pressure increase in the wave heating region is large enough to accelerate material upward and out of the heating region at supersonic velocities (exceeding $10^3 \, {\rm km}/{\rm s}$, see \autoref{fig:StructureR15Msun}) such that a cooling timescale by advection becomes shorter than a local dynamical timescale, limiting the buildup of pressure. This material decelerates when it runs into the massive overlying envelope.

\begin{figure}
\begin{center}
\includegraphics[scale=0.33]{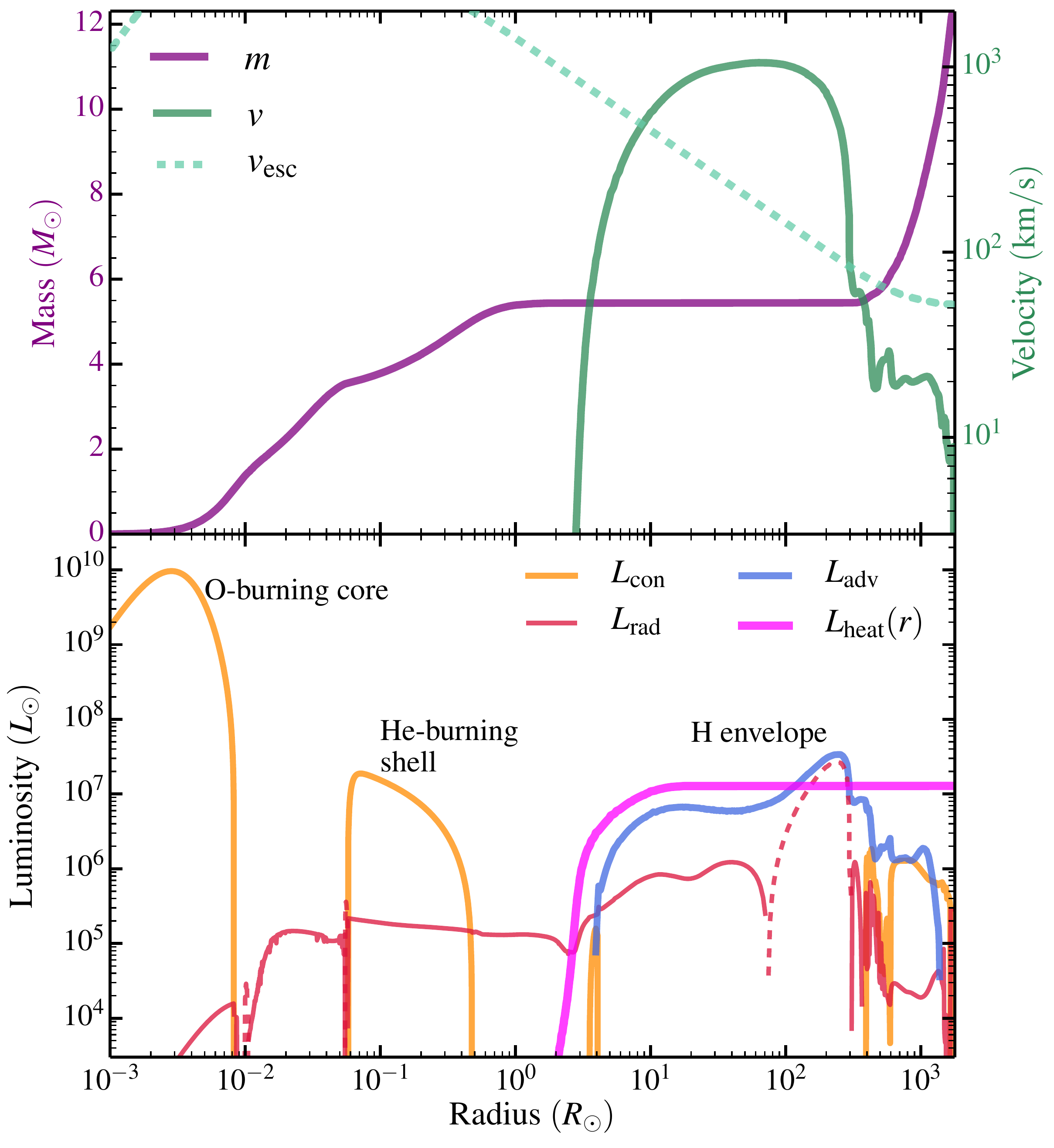}
\end{center} 
\caption{ \label{fig:StructureR15Msun} 
{\bf Top:} Interior mass and velocity as a function of radial coordinate in our model during core oxygen burning. The wave heating drives a wind that inflates a bubble of high velocity, low density material between the helium core and the overlying hydrogen envelope. Note the significant radial extent but tiny amount of mass within this evacuated bubble. The high velocity flows are contained by the massive overlying hydrogen envelope, a structure which will be modified by multi-dimensional instabilities (\autoref{RTI}). {\bf Bottom:} Convective, radiative, and advective energy fluxes in our model, with dashed lines indicating a negative (inward) energy flux. The magenta line is the integrated wave heating rate $L_{\rm heat}(r)$ out to radius $r$ (same as \autoref{fig:WaveDamp15Msun} but now plotted as function of radial coordinate).}
\end{figure}

As mass is accelerated out of the heating region, a peculiar structure develops: a dense helium core surrounded by an evacuated cavity filled by the low density wind, contained by a higher density but nearly stationary overlying envelope (\autoref{fig:StructureR15Msun}). In essence, the wave heating blows a nearly empty bubble at the base of the hydrogen envelope. As material is blown out of the heating region, it is replaced by upwelling material from beneath. The heating region digs down toward the helium core, and the mass coordinate of the base of the heating region decreases with time. Consequently, wave heat is distributed over a larger amount of mass ($\sim \! 10^{-2} \, M_\odot$ in our models) than it would be otherwise. The effective heating time (integrated over all mass that has absorbed wave energy) increases, becoming smaller than a dynamical time. For this reason, no strong pressure wave is driven into the envelope. Instead, the bubble inflates slowly, lifting the overlying envelope nearly hydrostaticly.

We caution that multi-dimensional effects are likely to drastically alter this scenario and the resulting density profile of the star, which we discuss further in \autoref{disc}. Nonetheless, the density structure of the RSG may be substantially altered by wave heating, with likely  implications for the lightcurve of its subsequent SN.

\section{Discussion}
\label{disc}

\subsection{Implications for Subsequent Supernovae}
\label{implications}

Our results have significant implications for SNe resulting from RSGs affected by wave heating. We have shown that waves can deposit $\sim \! 10^{48} \, {\rm erg}$ of energy into the stellar envelope (an amount comparable to its binding energy) in the last months to years of the star's life. Because this energy is negligible compared to the core binding energy, wave heating is unlikely to greatly alter the core structure or SN explosion mechanics (also, neutrinos can cool wave heated regions in the core).

\begin{figure}
\begin{center}
\includegraphics[scale=0.36]{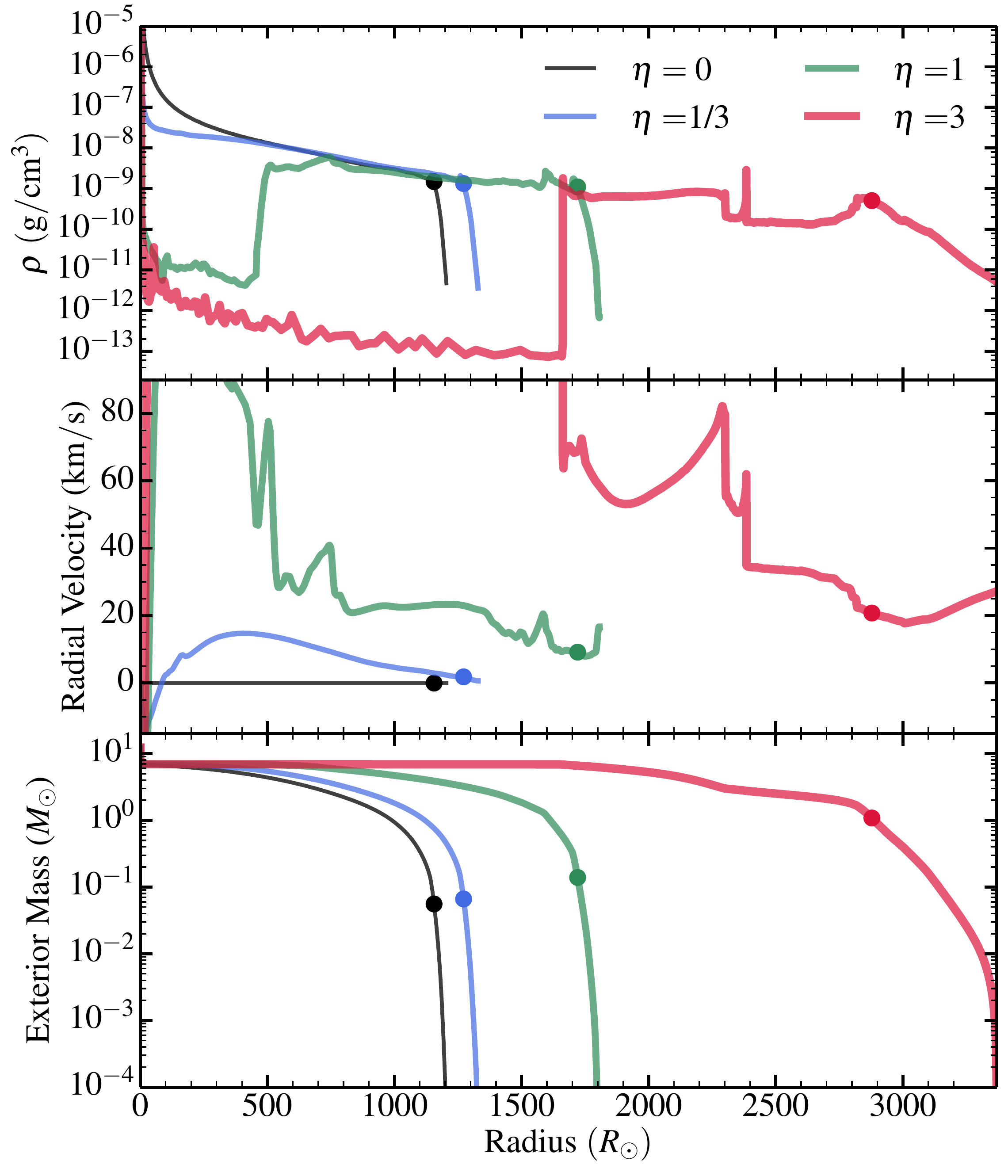}
\end{center} 
\caption{ \label{fig:StructureComp15Msun} 
{\bf Top:} Density profiles of our $M_{\rm ZAMS} = 15 \, M_\odot$ models during core oxygen burning, for different wave heating efficiencies. Dots are the location of the photosphere where $\tau = 2/3$. Stronger wave heating inflates larger (and lower density) bubbles beneath the hydrogen envelope, but Rayleigh-Taylor instabilities will likely smooth out much of this structure (\autoref{RTI}). {\bf Middle:} Corresponding radial velocity profiles. {\bf Bottom:} Exterior masses for the same models. Stronger heating ejects more mass into a circumstellar wind at higher velocities, and to greater distances above the photosphere. }
\end{figure}

The effect on the envelope structure, however, may be dramatic. The first crucial event in our models is the pressure wave breakout that results from wave heating during core Ne burning. For our nominal wave heating efficiency, a small amount of mass ($\sim \! 10^{-1} \, M_\odot$) is ejected at roughly one half the escape speed (see \autoref{fig:StructureComp15Msun}). Much of this mass falls back toward the star before core-collapse, and the resulting surface structure is neither hydrostatic nor does it have a steady wind density profile. However, we also note that several physical effects in the outflowing envelope material (e.g., treatment of convection, radiative transfer, non-spherical shock fronts, line-driven winds, molecule/dust formation) have not been properly treated in our models, and it is possible the outflow could have a component with somewhat higher velocity that extends to larger radii. For our optimistic wave heating efficiency ($\eta=3$), the outburst is strong enough to eject $\sim 1 \, M_\odot$ at $v \! \sim \! v_{\rm esc}$, producing a dense outflow up to the moment of core-collapse. Nominal outflow velocities of $\sim \! 30 \, {\rm km}/{\rm s}$ and timescales of $\sim \! 1 \, {\rm yr}$ imply the CSM is confined within $\sim \! 10^{14} \, {\rm cm}$ of the progenitor photosphere at the time of core-collapse.

The second crucial event occurs during core O-burning. In our models, O-burning inflates an evacuated bubble at the base of the H-envelope that lifts the overlying envelope to larger radii. The density structure of the envelope is substantially altered. The main effects (when plotting density vs. mass coordinate, see \autoref{fig:WaveDamp15Msun}) are to increase the envelope volume and decrease its density, and to flatten the density profile of the envelope. 

The wave-induced mass ejection events could substantially alter early SN spectra, and are a very compelling mechanism to produce the growing class of flash-ionized Type II-P/L SNe \citep{khazov:16,yaron:17} which show recombination lines from CSM at early times. The wave model predicts large (but not extreme) mass loss rates of $10^{-3} \, M_\odot/{\rm yr} \!  \lesssim \! \dot{M} \! \lesssim \! 10^0 \, M_\odot/{\rm yr}$, and slow velocities of $v \! \lesssim \! 100 \, {\rm km}/{\rm s}$ similar to those that have been measured or inferred. Crucially, the wave model explains why outbursts occur in the last months or years of the progenitor's life, which also accounts for the confinement of the CSM to small distances from the progenitor.

The altered density structure will also affect the SNe lightcurve. Shock cooling from a dense wind could create a faster rise time \cite{gonzalez-gaitan:15} that may alleviate the tension between measured galactic RSG radii and the suprisingly small radii inferred from shock cooling models without a wind \citep{gall:15,rubin:16}. The dense wind can also create early peaks in type IIP \cite{moriya:11,morozova:16}, and can cause the SNe to appear more IIL-like \citep{moriya:12}. Our optimistic wave efficiency produces CSM masses and density profiles similar to those inferred by \cite{morozova:16}, although our nominal wave efficiency does not appear to eject mass in a wind-like density profile due to mass fallback. Additionally, the flatter density profile of our models relative to non-heated models (see \autoref{fig:WaveDamp15Msun}) will result in a more steeply declining lightcurve \citep{pejcha:15}, again making the SN more II-L-like. We speculate that the altered density profile contributes substantially to the observed diversity of type II-P/II-L lightcurves, but more sophisticated SN light curve modeling will be needed for detailed predictions.

Supernova shock breakout could appear different from prior expectations in the presence of wave-induced mass ejection. In contrast to the steep density profiles near the photospheres of stellar models, detected shock breakouts \cite{schawinski:08,gezari:15} appear to emerge from a more extended photosphere or wind with a shallower density profile. Wave-induced mass loss can produce this sort of density structure (\autoref{fig:StructureComp15Msun}). However, even in the absence of wave heating, significant ``coronal" material may exist at the base of the wind-launching region \citep{dessart:17,moriya:17} and may also contribute to extended UV shock breakout and the optical SNe features discussed above.

Finally, it is unlikely that wave heating in ``normal" RSGs will lead to luminous type IIn SNe. The main reason is that there is not enough time to eject material to large radii of $\sim10^{15}-10^{16} \, {\rm cm}$ needed for a luminous IIn event. Even optimistic ejection speeds of $10^{7} \, {\rm cm}/{\rm s}$ and durations of $10^{8} \, {\rm s}$ cannot quite propel material to large enough distances (but see \autoref{binaries}).

\subsection{Comparison with Existing Observations}
\label{observations}

It is well established from pre-SN imaging that most type II-P SNe arise from RSG progenitors with inferred masses $M \! \lesssim \! 20 \, M_\odot$ \citep{smartt:09,smartt:09b,vandyk:12b,maund:14}. In many cases, progenitor characteristics have been measured from archival ground-based or {\it Hubble Space Telescope} data that predates the SN by more than $\sim \! 10$ years. In such  cases, we do not expect wave heating to significantly impact the appearance of the progenitor or its inferred mass. However, we encourage caution when inferring progenitor masses from pre-SN imaging. Our models predict that progenitors could be more luminous than expected, causing masses to be overestimated, at least when pre-SN imaging occurs after the onset of Ne/O burning. 

In a few cases (e.g., SN2003gd, \citealt{smartt:04}; SN2004A, \citealt{maund:14}, SN2008bk, \citealt{vandyk:12b}; ASASSN-16fq, \citealt{kochanek:17};), pre-explosion imaging was obtained within a few years of explosion. In most of these cases the SN progenitor was faint ($L \! < \! 10^5 \, L_\odot$), and the inferred mass was low ($M \! \lesssim \! 11 \, M_\odot$), significantly smaller than the $15 \, M_\odot$ model explored here (the inference of $M \! \sim \! 17 \, M_\odot$ for the progenitor of SN2012aw by \citealt{vandyk:12,fraser:12} has since been revised downward to $\approx \! 12 \, M_\odot$, see \citealt{kochanek:12}). Note also the convective overshoot in our model made it behave like a slightly more massive star of $\approx \! 17 \, M_\odot$, compared with other stellar evolution codes with less internal mixing. Future modeling of low-mass RSG SN progenitors will be needed to determine whether wave heating can strongly affect their pre-SN properties. SN2004A was imaged roughly 3 years before the SN, and was significantly brighter (and slightly cooler) than some of the other progenitors, possibly arising from a higher mass star \citep{maund:14}. We suggest the pre-SN properties of this star may have been affected by wave heating.

Multi-epoch photometry of the progenitor of ASASSN-16fq disfavors significant variability like that predicted in  \autoref{effects} \citep{kochanek:17}. The progenitor was estimated to be low-mass ($8 \, M_\odot \lesssim M \lesssim 12 \, M_\odot$), again significantly less massive than our model. These observations indicate that wave heating effects in that star were smaller than we have predicted for our higher mass model, or that pre-SN variability/outbursts only occur in a subset of type II-P progenitors. Preliminary wave heating calculations indicate that pre-SN variability may be smaller in progenitors with $M \! \sim \! 10 \, M_\odot$ due to longer evolution timescales and lower wave heating rates. Future work examining wave heating in lower mass RSG progenitors will be necessary for detailed observational comparisons.


\subsection{Predictions}
\label{predictions}

The strongest prediction of our work is that mild pre-SN outbursts will be common in RSG progenitors of type II SNe. Although we have not explored the entire parameter space of RSG masses and properties, our otherwise ``normal" model suggests similar effects to those explored here will operate in many RSGs. In lower mass RSGs, there may be multiple smaller amplitude outbursts spread over the last $\sim$decade of the star's life due to multiple core burning phases. Higher mass RSGs are expected to exhibit fewer but larger amplitude outbursts, occurring in the final $\sim$months of life.

We also predict that most RSG outbursts will exhibit modest luminosity excursions of less than $\sim$2 magnitudes. We expect peak bolometric luminosities to remain under $\sim \! 10^{6} \, L_\odot$. Ejecta masses will likely be small, $M_{\rm ej} \lesssim 1 \, M_\odot$, and with low velocities $v \lesssim 50 \, {\rm km}/{\rm s}$. These mild outbursts will be missed by most current transient surveys, but upcoming surveys with greater sensitivity and higher cadence (e.g., ZTF, BlackGem,  LSST) may verify or rule out our predictions.

We predict wave heating to increase the luminosity of the resulting SN due to the inflated progenitor radius. Analytic scalings predict plateau luminosities of \citep{popov:93,kasen:09,sukhbold:16}
\beq
L_p \propto E_{\rm SN}^{5/6} M_{\rm env}^{-1/2} R^{2/3} \,
\eeq
and plateau durations
\beq
t_p \propto E_{\rm SN}^{-1/6} M_{\rm env}^{1/2} R^{1/6} \, ,
\eeq
where $E_{\rm SN}$ is the SN explosion energy, $M_{\rm env}$ is the envelope mass, and $R$ is the pre-SN stellar radius. Hence, we expect the plateau duration to be insensitive to wave heating, but the SN luminosity may be significantly larger ($L \! \propto \! R^{2/3}$) for the same explosion energy. Alternatively, the larger progenitor radii (by a factor of $\sim \! 2$) of our models would require smaller explosion energies, all else being equal.

\subsection{Rayleigh-Taylor Instabilities}
\label{RTI}

The density profiles shown in \autoref{fig:StructureComp15Msun} are unrealistic because of multi-dimensional effects, in particular because of the Rayleigh-Taylor instabilities (RTI) that will exist real stars. RTI can operate when pressure and density gradients have the opposite sign \citep{chandrasekhar:61,duffell:16}, for instance, in massive star atmospheres where density inversions predicted by 1D models are altered by RTI \citep{jiang:15,jiang:16}. In our case, RTI will occur at the surface of the wind-blown bubble during O-burning. The interface between the inflated cavity (high pressure, low density) and overlying envelope (low pressure, high density) will give rise to RTI which will likely act to smooth the density profiles shown in \autoref{fig:StructureComp15Msun}. The mixing produced by RTI may allow more envelope material to mix downward into the heating region, and allow more heated material to mix upward into the envelope. The net effect on the RSG envelope structure is unclear, but the very large and low density cavities in \autoref{fig:StructureComp15Msun} will likely shrink and increase in density. Nonetheless, the envelope density profile may be strongly altered by wave heating during O-burning.

\subsection{Caveats}
\label{caveats}

Because this is one of the first investigations of the hydrodynamic/observational details of wave-driven heating, there are a number of uncertainties and caveats that must be considered.

\subsubsection{Wave Excitation}

Probably the largest uncertainty in our calculations is the amplitude and spectrum of gravity waves excited by convection in nuclear burning zones. We have approximated the gravity waves as monochromatic in both temporal and horizontal wavenumber (one frequency and spherical harmonic index $\ell$) which is clearly a gross simplification. If the waves are excited to lower amplitudes (e.g., because we have calculated $L_{\rm wave}$ at an inappropriate location) or higher amplitudes (e.g., because wave luminosity scales as $\mathcal{M}_{\rm con}^{5/8}$ as suggested by \citealt{lecoanet:13}), the wave heating effects will be significantly altered, as demonstrated by the reduced and enhanced wave efficiency factors $\eta$ in Figures \ref{fig:Edep15MsunComp}, \ref{fig:HR15MsunComp}, and \ref{fig:StructureComp15Msun}. The wave frequency spectrum excited by convection is not well understood, as \cite{goldreich:90,lecoanet:13} argue for excitation at $\omega_{\rm wave} \! \sim \! \omega_{\rm con}$ due to bulk Reynolds stresses, while \cite{rogers:13} argues for excitation via plume incursion that adds a substantial high frequency ($\omega_{\rm wave} \! > \! \omega_{\rm con}$) tail to the spectrum. If our estimates of wave frequencies are too high/low, then we have likely over/underestimated the fraction of wave energy that heats the envelope ($f_{\rm esc}$, equation \ref{waveheat}) because high/low frequency waves are less/more subject to neutrino damping and usually have a higher/lower transmission coefficient into the envelope (see \autoref{waveprop}). Finally, if waves are mostly excited at higher angular wavenumbers than $\ell=1$, heating rates will be substantially reduced because higher angular wavenumbers are more strongly damped and have smaller transmission coefficients.

\subsubsection{Nonlinear Effects}

All calculations in this work assume wave amplitudes are small enough for linear wave physics to apply, which may be reasonable where $k_r \xi_r \gtrsim 1$, with $k_r$ the radial wavenumber and $\xi_r$ the radial displacement. Preliminary waveform solutions indicate this criterion is satisfied for high frequency waves with $\omega \gtrsim 2 \omega_{\rm con}$, but not for lower frequency waves. These waves may be attenuated by non-linear wave breaking in the core, so if the wave power spectrum contains most of its power at frequencies less than $\sim 2 \omega_{\rm con}$, our wave heating rates will be significantly overestimated. We intend to investigate this more thoroughly in a future publication. Additionally, non-linear coupling and instabilities are known to operate at smaller amplitudes (see e.g., \citealt{weinberg:08}) in various contexts. If non-linear coupling in the g mode cavity is able to prevent waves from being transmitted into the envelope, this further could suppress wave heating.

\subsubsection{Convection and Radiative Transfer}

Our one dimensional simulations implement MLT for convective energy transport, and the diffusion approximation for radiative energy transport. The former approximation is calibrated for stars in hydrostatic and thermal equilibrium, which is not the case in the outflowing near-Eddington envelope of our models. In the wave heating region, we have utilized acceleration-limited convective velocities (see \autoref{models}), with a maximum acceleration of the mixing length velocity equal to the local gravitational acceleration, $g$. However, if it can accelerate faster, convection at the base of the hydrogen envelope could carry more wave heat outward because the maximum convective luminosity $L_{\rm max} = 2 \pi r^2 \rho c_s^3 \gg L_{\rm heat}$ in this region. We have performed experiments without limiting convective acceleration, finding the pressure wave launched during Ne burning and the final stellar radius are only are weakly affected. However, during O burning, convection carries most of the wave energy outward from the heating region, causing the star to reach much higher surface luminosities of $\gtrsim \! 10^6 \, L_\odot$. The wave-inflated cavity still exists but is smaller and less evacuated. A better understanding of convection's ability to respond to sudden heating is needed for robust predictions of the stellar luminosity and density evolution.

In addition to affecting the background envelope structure, the use of MLT will affect the luminosity during the pressure wave breakout. It is not immediately clear how to treat convective energy transport in the regime where bulk velocities are a significant fraction of the sound speed.  We have experimented with different treatments of convection (e.g., limiting maximum convective velocities), finding they produce modest quantitative alterations of our results but do not change the basic picture. The use of the diffusion approximation may also produce errors in our predicted pressure wave breakout luminosity evolution, which we hope to re-examine in future work.

\subsubsection{Rotation and Flows}

We have ignored effects of rotation in this preliminary analysis, which is justified in the slowly rotating stellar envelope. Rotation could significantly affect wave excitation and propagation in the core if its rotation rate is comparable to wave angular frequencies, but late stage core rotation rates are poorly constrained. Rapid core rotation will probably not eliminate wave heating because it is difficult to suppress both prograde and retrograde waves with reasonable rotation profiles, although the wave heating efficiency could be reduced.

In this work we did not include background flows in equations governing wave propagation, even though we showed that waves can generate supersonic flows within the stellar envelope. Our approximation is valid during core Ne burning when induced velocities are small compared to wave group velocities. During core O burning, however, some wave energy damps in regions where flow velocities are comparable to the sound speed (e.g., near $10 \, R_\odot$ in \autoref{fig:StructureR15Msun}). Such flows will alter wave propagation/dissipation, but we leave this for future work in light of the additional effects of shock formation and Rayleigh-Taylor instabilities that will also alter flow velocities (see below).


\subsection{Magnetic Fields}

Background magnetic fields may be important in some stars. We do not expect them to greatly alter the envelope dynamics where the waves are acoustic in nature and the flow velocities are mostly radial. However, sufficiently strong magnetic fields can prevent gravity wave propagation in the core \citep{fuller:15}. Such fields would likely confine wave energy to the core of the star and prevent wave heating outbursts. We discuss this possibility in \autoref{magnetic}.

\subsection{Binaries}
\label{binaries}

Binary interactions may contribute to pre-SN mass loss \citep{chevalier:12} but need to be finely tuned to occur in the final years of evolution. It might be possible, however, for the combination of wave heating and binary interactions to produce IIn SNe in a small fraction of RSGs. If the RSG has been partially stripped of its H-envelope, wave heat will be concentrated in a smaller amount of mass and larger ejection speeds may be possible. Furthermore, outburst luminosities in stripped stars will be much larger due the smaller thermal time of the envelope (Fuller 2017, in prep). Finally, envelope inflation via waves could induce a common envelope event for an appropriately placed binary companion, potentially ejecting more mass at larger speeds and creating a IIn event \citep{mcley:14}.

\subsection{Relation to other Theories of Pre-SN Outbursts}
\label{other}

The notable feature of wave-driven outbursts is its generality: it can occur in low-mass ($M \! \! < \! \! \, 20 \, M_\odot$) stars that are the most common SNe progenitors. Below, we discuss other mass-loss mechanisms that have been proposed, but note that many are restricted to small regions of SN progenitor parameter space or do not yet yield quantitative predictions.

One possible mechanism for pre-SN outbursts is instabilities during late stage (C/Ne/O) convective shell burning. In a series of papers \citep{meakin:06,meakina:07,meakinb:07,Arnett_2008,arnett:11,smitharnett:14,cristini:16}, Meakin, Arnett, and collaborators have investigated the properties of convection during late phase (carbon shell burning and beyond) nuclear burning. They find that the convective burning shells exhibit some interesting properties not predicted by mixing length theory (therefore not typically implemented in 1D stellar evolution codes), such as entrainment and energy generation rate fluctuations. However, it remains unknown whether convective fluctuations can grow large enough to produce any detectable effect at the stellar surface, nor is it clear what the observational signature would be and how often this process should occur.

\cite{murphy:04} examined linear instabilities during late burning phases, finding no instabilities growing fast enough to produce large effects. \cite{woosleyheger:15} showed that degenerate Si-burning flashes in $\simeq \! 10 \, M_\odot$ stars could produce shock waves that eject part of the stellar envelope, which may account for some fraction of IIn SNe. Additionally, pair instabilities in very massive stars ($M\gtrsim 60 \, M_\odot$) may produce some outbursts and interacting SNe \citep{woosley:16}, but again the rarity of these events and distinct light curve features makes them unlikely to be responsible for most type IIn SNe. \cite{heger:97} and \cite{yooncantiello:10} show that envelope pulsational growth rates increase after core helium depletion, potentially driving a superwind during the last tens of thousands of years of a star's life, although this theory cannot explain very high ($ > \! 10^{-3}\, M_\odot/{\rm yr}$) mass loss rates in the last years of a star's life (except perhaps in very massive stars, \citealt{moriya:15}). \cite{soker:17} suggest intense core dynamo activity can generate outbursts through the buoyant rise of magnetic flux tubes, but they neglect to account for stable stratification in radiative shells which can strongly hinder the radial motion of flux tubes and prevent them from rising into the envelope. Mass loss can be triggered by the loss of gravitational binding energy due to neutrino emission \citep{moriya:14}, but this can only occur for stars extremely close to the Eddington limit and can only yield $ \dot{M}> \! 10^{-3}\, M_\odot/{\rm yr}$ during the last $\sim$month of the star's life.

\section{Conclusions}
\label{conclusions}

We have modeled the evolution of a $15 \, M_\odot$ red supergiant (RSG) model in the final decades before core-collapse, accounting for energy transport by convectively excited waves. Our goal was to determine whether wave energy transport can affect the pre-supernova (SN) structure of the star or produce pre-SN outbursts as suggested by \cite{quataert:12,shiode:14}. We used the MESA stellar evolution code \citep{paxton:11,paxton:13,paxton:15} to model the effect of wave heating on the stellar structure, implementing its 1D hydrodynamical capabilities to capture shocks and outflows resulting from wave heating.

During late nuclear burning phases (core Ne and O burning in particular), convective luminosities of $L_{\rm con} \! \sim \! 10^{10} \, L_\odot$ will excite gravity waves which carry energy fluxes of $L_{\rm wave} \! \sim \! 2 \! \times \! 10^{7} \, L_\odot$. We calculate that much of this energy will be transmitted into acoustic waves that propagate out of the core and into the envelope, carrying a flux of $L_{\rm heat} \! \! \sim \! \! 10^7 \, L_\odot$. The acoustic waves damp into thermal energy near the base of the hydrogen envelope due to the large drop in density at that location. In our models, wave heating during core Ne burning launches a pressure wave that propagates toward the stellar surface, steepening into a weak shock that creates a mild outburst $\sim \! \! 1 \, {\rm yr}$ before core-collapse. The outburst is dim by SN standards ($L \! \sim \! 3 \times 10^5 \, L_\odot$, \autoref{fig:HR15MsunComp}), and ejects a small amount of mass ($M_{\rm ej} \lesssim 1 \, M_\odot$) at low velocities ($v \lesssim 50 \, {\rm km}/{\rm s}$, \autoref{fig:StructureComp15Msun}).

In our models, wave heating during core O burning drives a wind off the surface of the He core, inflating a low density bubble that gradually lifts off the overlying H envelope. However, we expect Rayleigh-Taylor instabilities to strongly modify these dynamics, potentially leading to another outburst during O burning. Regardless, the H envelope can be significantly inflated, with a non-hydrostatic density profile differing from prior expectations. 

We do not expect wave heating to lead to very luminous type IIn SNe in ``normal" $M \! \lesssim \! 20 \, M_\odot$ RSG progenitors because the modest amount of ejected mass is confined at small distances ($\lesssim \! 10^{15} \, {\rm cm}$) from the RSG. However, we find wave heating is a compelling mechanism to produce flash ionized type II-P/II-L SNe (e.g., \citealt{khazov:16,yaron:17}) showing emission lines in early spectra. The altered density structure will affect the resulting SN luminosity, potentially producing an early peak or a more II-L-like light curve, contributing to the diversity of type II SNe.

The physics of wave-driven outbursts is rich, involving complex hydrodynamic processes spanning nearly 20 orders of magnitude in density. Our results are thus subject to numerous caveats discussed in \autoref{caveats} that can be improved with future work. It will also be necessary to examine wave heating in other SN progenitors (e.g., different stellar masses, metallicities, rotation rates, binarity, degree of envelope stripping, etc.) to understand how wave-driven outbursts contribute to the enormous diversity of core-collapse SNe.

\section{Acknowledgments}

We thank Matteo Cantiello, Bill Paxton, Stephen Ro, Maria Drout, Nathan Smith, Schuyler Van Dyk, Jeremiah Murphy, Eliot Quataert, and Lars Bildsten for useful discussions. JF acknowledges partial support from NSF under grant no. AST-1205732 and through a Lee DuBridge Fellowship at Caltech. This research was supported in part by the National Science Foundation under Grant No. NSF PHY-1125915, and by the Gordon and Betty Moore Foundation through Grant GBMF5076.

\bibliography{MassiveWavesBib}

\appendix

\section{Massive Star Models with MESA}
\label{models}

\subsection{Evolving to Carbon Burning}

We created stellar models using the MESA stellar evolution code \citep{paxton:11,paxton:13,paxton:15}, version 9393. Our model evolution proceeded in three steps. First, we evolved a $15 \, M_\odot$ model from the main sequence to just before the onset of core carbon burning. Most model settings are default values, and the models are non-rotating with $Z = 0.02$.

One notable change is to add a significant amount of overshooting to our models via the inlist setting
\begin{verbatim}
overshoot_f_above_nonburn_core = 0.025
overshoot_f0_above_nonburn_core = 0.01
\end{verbatim}
and using the same overshoot/undershoot values for H,He, and Z core/shell burning. This corresponds to an exponential overshoot parameter of $f_{\rm ov} \simeq 0.015$. We use the following mass-loss prescription settings:
\begin{verbatim}
hot_wind_scheme = 'Dutch'
cool_wind_RGB_scheme = 'Dutch'
cool_wind_AGB_scheme = 'Dutch'
RGB_to_AGB_wind_switch = 1d-4
Dutch_scaling_factor = 0.8
\end{verbatim}
This model has He core mass $M_{\rm He} = 5.38 \, M_\odot$ and total mass $M = 12.31 \, M_\odot$ at the onset of carbon burning. The helium core mass is somewhat larger than models not including overshoot, and make our model behave like a slightly more massive star compared to some other stellar evolution codes.

We add a small amount of element diffusion (comparable to what has been asteroseismicly inferrred, \citealt{moravveji:15}) to our models to slightly smooth sudden composition/density jumps, which produce large (possibly unphysical) spikes in the Brunt-V\"{a}is\"{a}l\"{a} frequency $N$, using
\begin{verbatim}
set_min_D_mix = .true.
min_D_mix = 1d2
\end{verbatim}
Additionally, we restrict changes in composition at each timestep due to nuclear burning with
\begin{verbatim}
dX_div_X_limit_min_X = 1d-5
dX_div_X_limit = 1d-1
dX_nuc_drop_min_X_limit = 3d-5
dX_nuc_drop_limit = 3d-3
\end{verbatim}
which helps ensure more accurate composition profiles as nuclear burning processes begin and end within the core. This helps prevent the occurrence of, e.g., unphysical violent burning flashes when Ne ignites due to residual unburnt carbon.

We add wave heating (described below) throughout the entire evolution, however the wave energy is totally negligible (orders of magnitude below the surface luminosity) at all points proceeding carbon burning.

\subsection{Preparing for Hydrodynamics}

Before the onset of carbon burning, we save a model as our basepoint for the evolutions presented in this paper. We then load and run this model, with the following \verb|star_job| command: 
\begin{verbatim}
relax_initial_tau_factor=.true. 
relax_to_this_tau_factor=1d-4
dlogtau_factor=.1
\end{verbatim}
which allows the model to evolve material above the photosphere out to an optical depth $\tau = 10^{-4}$. After relaxation, we evolve the model with a maximum timestep of one year for twenty-five models, the small timestep assuring the model is very close to hydrostatic equilibrium.

\subsection{Running with Hydrodynamics}

After relaxing our model, we turn on the hydrodynamics capabilities of MESA with
\begin{verbatim}
change_initial_v_flag = .true.
change_v_flag = .true.
new_v_flag = .true.
\end{verbatim}
This introduces a very small transient in surface temperature and luminosity, but we caution that a non-relaxed model may exhibit much larger transients and struggle converge when hydrodynamics are first turned on. 

At the outer boundary of our model, we let mass flow outward by removing it below a density of $\rho_{\rm min} = 2 \times 10^{-14} \, {\rm g}/{\rm cm}^3$ to avoid equation of state problems for matter at lower densities
\begin{verbatim}
remove_surface_by_density = 2d-14
repeat_remove_surface_for_each_step = .true.
\end{verbatim}
although none of our models actually reach outer boundary densities this small. 

We use the following settings to limit the convective energy transport via MLT in MESA:
\begin{verbatim}
mlt_accel_g_theta = 1
min_T_for_acceleration_limited_conv_velocity=0d0
max_T_for_acceleration_limited_conv_velocity=1d11
max_conv_vel_div_csound = 1d0
\end{verbatim}
The first three commands limit the changes in convective velocities/fluxes due to sudden developments of temperature gradients, e.g., in the wave heating region or near shocks. Failure to limit convective velocities will allow convection to transport energy toward the surface and across shocks at unphysically large rates. This prescription may not be optimal, but is more realistic than allowing instantaneous increases in convective fluxes.

The following commands control the hydro equations and boundary conditions solved at each timestep
\begin{verbatim}
use_ODE_var_eqn_pairing=.true.
use_dvdt_form_of_momentum_eqn=.true.
use_dPrad_dm_form_of_T_gradient_eqn=.true.
use_compression_outer_BC=.true.
use_T_Paczynski_outer_BC = .true.
\end{verbatim}
We find these outer boundary conditions to be fairly stable. Experiments with other boundary conditions appear to produce similar results, but are much more likely to cause the code to crash or to produce unphysical jumps in surface temperature, especially when a shock is propagating near the photosphere.

Spatial gridding and error tolerances are adjusted with the following controls
\begin{verbatim}
okay_to_remesh = .true.
min_dq=1d-14
log_tau_function_weight=50
log_kap_function_weight=50
R_function_weight = 50
newton_iterations_limit=9
iter_for_resid_tol2=6
tol_residual_norm1=1d-8
tol_max_residual1=1d-7
tiny_corr_coeff_limit=999999
newton_itermin_until_reduce_min_corr_coeff=999999
\end{verbatim}
It is necessary to adjust the grid weights, otherwise very low density regions above the photosphere and within the empty cavity during O-burning are not well-resolved.

During core O-burning, an instability develops within the supersonic wind at the base of the H-envelope. The instability appears to stem from the sonic point of the flow, such that the flow below the sonic point is smooth, but large velocity/density inhomogeneities develop above. Although radial and nonradial instabilities may exist \citep{shaviv:99,shaviv:01}, we believe the instability in MESA is a numerical artifact, because it is largely suppressed in the absence of convection. In our runs, we prevent convection at this sonic point by adding the following command to MESA's MLT module:
\begin{verbatim}
if ((abs(s% v_start(k))) >= 5d6) 
   max_conv_vel = 0d0
end if
\end{verbatim}
which prevents convection in regions with velocities larger than $50 \, {\rm km}/{\rm s}$. Convection can still operate near the surface where velocities are typically smaller than this limit. We have performed simulations with and without this fix, and it does not appear to strongly affect the development of the wind, except that using the fix prevents the formation of internal shocks within the wind and allows the code to run much faster. We defer a more detailed analysis because the entire wind configuration will likely be altered by RTI as discussed in \autoref{RTI}.

Finally, we add a small amount of numerical viscosity beginning during O-burning (after the Ne pressure wave breakout):
\begin{verbatim}
viscosity_factor = 1d-4
\end{verbatim}
This helps the code run faster in the presence of strong shocks that can develop at interfaces between the wave-driven wind and overlying envelope.

\section{Wave Propagation}
\label{waveprop}

Here we derive the fraction of wave energy which is able to tunnel into the envelope and dissipate into thermal energy.

\subsection{Wave Damping via Neutrinos}

The wave entropy perturbation per unit mass due to neutrinos is \citep{unno:89}
\beq
\label{dsneut}
 i \omega T \delta S_\nu = \epsilon_\nu \bigg[ \bigg( \frac{\partial \ln \epsilon_\nu}{\partial \ln T} \bigg)_{\!\!\rho} \frac{\delta T}{T} + \bigg( \frac{\partial \ln \epsilon_\nu}{\partial \ln \rho} \bigg)_{\!\!T} \frac{\delta \rho}{\rho} \bigg] \, .
\eeq
Here, $\epsilon_\nu$ is the neutrino cooling rate per unit mass, the terms in parentheses are its partial derivatives with respect to temperature and density, and $\delta T$ and $\delta \rho$ are the Lagrangian perturbations in temperature and density produced by the wave. The energy loss rate (when integrating over a wave cycle) per unit mass is then
\begin{align}
\label{deneut}
\delta \epsilon_{\nu} &= \delta T \frac{d \delta S}{d t} \nonumber \\
&= \epsilon_\nu \frac{\delta T}{T} \bigg[ \bigg( \frac{\partial \ln \epsilon_\nu}{\partial \ln T} \bigg)_{\!\!\rho} \frac{\delta T}{T} + \bigg( \frac{\partial \ln \epsilon_\nu}{\partial \ln \rho} \bigg)_{\!\!T} \frac{\delta \rho}{\rho} \bigg] \, .
\end{align}

Now, in the nearly adiabatic limit of interest, the temperature perturbation is 
\beq
\label{dt}
\frac{\delta T}{T} = \frac{ \Gamma_1 \nabla_{\rm ad}}{c_s^2} \big( r \omega^2 \xi_\perp - g \xi_r \big) \, .
\eeq
where the thermodynamic quantities have their usual meaning, $\xi_r$ is the radial wave displacement, and $\xi_\perp$ is the horizontal displacement. Essentially all of the wave neutrino losses occur in the radiative core where the waves are well approximated as WKB gravity waves. For gravity waves, $\xi_r \sim \omega \xi_\perp/N \sim \omega c_s \xi_\perp/g$, and $\omega \ll c_s/r$. Therefore, the second term in equation \ref{dt} dominates, and 
\beq
\label{dtapprox}
\frac{\delta T}{T} \simeq \frac{\Gamma_1 \nabla_{\rm ad} g}{c_s^2} \xi_r \, .
\eeq
Additionally, neutrino loss rates are usually much more sensitive to temperature than density, so the first term in brackets in equation \ref{deneut} dominates. The energy loss rate via neutrinos is then
\beq
\label{deneu2}
\delta \epsilon_{\nu} \simeq \frac{\Gamma_1^2 \nabla_{\rm ad}^2 g^2}{N^2 c_s^4} \omega^2 \xi_\perp^2 \bigg(\frac{\partial \ln \epsilon_\nu}{\partial \ln T} \bigg)_{\!\!\rho} \epsilon_\nu .
\eeq
For gravity waves, the wave energy per unit mass is $\varepsilon \simeq \omega^2 \xi_\perp^2$. So the wave energy damping rate per unit time is 
\beq
\label{gamma_neu}
\gamma_{\nu} = \frac{ \delta \epsilon_{\nu}}{\varepsilon} \simeq \frac{\Gamma_1^2 \nabla_{\rm ad}^2 g^2}{N^2 c_s^4} \bigg(\frac{\partial \ln \epsilon_\nu}{\partial \ln T} \bigg)_{\!\!\rho} \epsilon_\nu .
\eeq

\subsection{Wave Tunneling into the Envelope}
\label{wavetunnel}

Calculating the wave energy flux tunneling into the envelope as acoustic waves is not straightforward because there may be multiple evanescent zones separating the generated waves from the envelope. Additionally, wave energy may damp out via neutrinos along the way. Thus, it is important to keep track of where wave energy builds up and how fast it damps out.

To calculate the amount of energy tunneling into the envelope, we can treat the star as a series of wave cavities separated by intervening evanescent regions. Within each wave cavity, the wave energy flux is conserved unless damping processes operate. At each evanescent region, only a fraction $T^2$ of the incident wave energy is able to tunnel through, where $T^2$ is the squared transmission coefficient of the evanescent region, which is approximately equal to \citep{unno:89}
\beq
\label{t2}
T_{1,2}^2 =  \exp \bigg( -2 \int^{r_2}_{r_1} |k_r| dr \bigg) \,  
\eeq
where $r_1$ and $r_2$ are the radial boundaries of the evanescent region, and the radial wavenumber is
\beq
\label{kr2}
k_r^2 = \frac{\big(N^2 - \omega_2 \big)\big(L_l^2 - \omega^2\big)}{\omega^2 c_s^2} \, .
\eeq
Note that $k_r$ is imaginary in evanescent zones.
In the limit of a thin evanescent region, equation \ref{t2} needs to be slightly modified \citep{takata:16}, although we shall see below that thick evanescent regions dominate the wave trapping.

In a steady state, the amount of energy entering and exiting each wave cavity is equal. The energy transfer rate from cavity 1 to cavity 2 through an evanescent region from $r_1$ to $r_2$ is 
\beq
\label{edot12}
\dot{E}_{1,2} = \frac{T_{1,2}^2}{2 t_1} E_1 \,
\eeq
where $E_1$ is the wave energy within cavity 1 and $t_1 = \int dr/v_g$ is the wave crossing time across cavity 1. Similarly, the energy transfer rate from cavity 2 to cavity 1 from $r_2$ to $r_1$ is
\beq
\label{edot21}
\dot{E}_{2,1} = \frac{T_{1,2}^2}{2 t_2} E_2 \,
\eeq
where we have used the fact that $T_{1,2}^2 = T_{2,1}^2$. The steady-state approximation is justified by the fact that the wave crossing timescales in the core of the star are typically much smaller than the nuclear burning timescales.

Consider the cavity (labeled as cavity 1) overlying the wave generation region, which has a wave energy input $L_{\rm wave}$. We will also consider damping processes within cavity 1 such that the energy loss to wave damping is $\dot{E}_{1,{\rm damp}} = E_1 \gamma_1$. Then balancing energy input and energy losses for cavity 1 yields
\beq
\label{edot1}
L_{\rm wave} + \dot{E}_{2,1} = \dot{E}_{1,2} + E_1 \gamma_1 \, .
\eeq
In our problem, neutrino damping is always largest closest to the wave generation site (cavity 1) where temperature and density are highest, so we ignore damping in overlying cavities. The net energy flux through overlying cavities is then $L_{\rm heat} = L_{\rm wave}-\dot{E}_{1,{\rm damp}}$, and our goal is to calculate $L_{\rm heat}$. The energy balance for cavity 2 is
\beq
\label{edot2}
L_{\rm wave} - E_1 \gamma_1 + \dot{E}_{3,2} = \dot{E}_{2,3} \, ,
\eeq
and a similar equation holds for overlying cavities. Rearranging equation \ref{edot2}, 
\beq
\label{edot2b}
\frac{E_2}{2t_2} = \frac{1}{T_{2,3}^2} \bigg[ L_{\rm heat} + \dot{E}_{3,2} \bigg] \, ,
\eeq
and substituting into equation \ref{edot1}, we have
\beq
\label{edot1b}
L_{\rm heat} + \frac{T_{1,2}^2}{T_{2,3}^2} \bigg[ L_{\rm heat} + \dot{E}_{3,2} \bigg] = \dot{E}_{1,2} \, .
\eeq
We can perform a similar procedure to substitute in for $\dot{E}_{3,2}$ and all overlying cavities up to cavity $n$, with the boundary condition of no wave flux entering from above, $\dot{E}_{n+1,n} = 0$. Then we have 
\beq
\label{edot1c}
L_{\rm heat} +  L_{\rm heat} T_{1,2}^2 \sum_2^n \frac{1}{T_{n,n+1}^2} = \dot{E}_{1,2} \, .
\eeq
Now, using $E_1 \gamma_1 = L_{\rm wave} - L_{\rm heat}$, we have
\beq
\label{edot1d}
L_{\rm heat} +  L_{\rm heat} T_{1,2}^2 \sum_2^n \frac{1}{T_{n,n+1}^2} = \frac{L_{\rm wave}-L_{\rm heat}}{2 \gamma_1 t_1} T_{1,2}^2  \, .
\eeq
which can be rewritten as
\beq
\label{lheat}
L_{\rm heat}  = L_{\rm wave} \bigg[ 1 + 2 \gamma_1 t_1 \sum_1^n T_{n,n+1}^{-2} \bigg]^{-1} \, .
\eeq

Equation \ref{lheat} is the desired result, it allows us to compute the wave energy escaping into the envelope, $L_{\rm heat}$ relative to the wave energy input rate $L_{\rm wave}$. All quantities on the right hand side can be computed from the stellar structure. Terms with large transmission coefficients ($T^2 \simeq 1$) should be replaced with the value $T^2 \rightarrow -\ln(1-T^2)$ \citep{takata:16}. However, terms with small values of $T^2$ dominate the sum in the right hand side of equation \ref{lheat}. In practice, the thickest evanescent zone usually dominates the sum, which can be well approximated by
\beq
\label{lheatapprox}
L_{\rm heat}  = L_{\rm wave} \bigg[ 1 + \frac{2 \gamma_1 t_1}{T_{\rm min}^{2}}\bigg]^{-1} \, ,
\eeq
where $T_{\rm min}^2$ is the minimum transmission coefficient between the side of wave generation in the core and wave dissipation in the envelope. In our models, this evanescent zone is usually created by the convective He burning shell.

The value of $\gamma_1$ accounts for damping throughout cavity 1. For neutrinos, the local damping rate is given by $\gamma_\nu$ in equation \ref{gamma_neu}. Upon traversing cavity 1, the wave energy is attenuated by a factor 
\begin{align}
\label{fneu}
f_\nu &= e^{x_\nu} = \exp \bigg[ 2 \int^{r_{1+}}_{r_{1-}} \frac{\gamma_\nu dr}{v_g} \bigg] \nonumber \\
&= \exp \bigg[2 \int^{r_{1+}}_{r_{1-}} \gamma_\nu \frac{\sqrt{l(l+1)}N dr}{\omega^2 r} \bigg] \, .
\end{align}
where $v_g \simeq \omega^2 r/(\sqrt{l(l+1)} N)$ is the radial group velocity of gravity waves, and $r_{1+}$ and $r_{1-}$ are the upper and lower boundaries of cavity 1. Then the time-averaged damping rate of the wave due to neutrino damping in cavity 1 is
\beq
\gamma_{1,\nu} = \frac{1-f_\nu^{-1}}{2 t_1} \simeq \frac{x_\nu}{2 t_1} \, .
\eeq
The second equality arises from the fact that in our models $x_\nu$ in equation \ref{fneu} is small, and $f_\nu \simeq 1 + x_\nu$.

Additional damping can occur during shell burning phases, when convectively excited waves tunnel into the radiative core. In this case, the wave amplitudes near the center of the star are large enough to induce non-linear wave breaking (see \citealt{fullerwave:15} and references therein). Thus, waves entering the central radiative region will be lost, which could occur if the waves excited from shell convection reflect from an overlying evanescent zone and then tunnel back through the burning shell and into the core. This effect can be modeled as an additional source of damping in cavity 1,
\beq
\label{gammatunnel}
\gamma_{1,{\rm core}} = \frac{T_{\rm shell}^2}{2 t_1} \, ,
\eeq
where $T_{\rm shell}^2$ is the transmission coefficient through the burning shell that excites the wave.

Accounting for both neutrino damping in cavity 1 and wave tunneling into the core, the effective damping rate in cavity 1 is $\gamma_1 = \gamma_{1,\nu} + \gamma_{1,{\rm core}}$. Using equation \ref{lheatapprox}, we arrive at our final expression determining the wave flux entering the envelope
\beq
\label{lheatfinal}
L_{\rm heat} = f_{\rm esc} L_{\rm wave} = \bigg[ 1 + \frac{T_{\rm shell}^2 + x_\nu}{T_{\rm min}^{2}} \bigg]^{-1} L_{\rm wave} \, ,
\eeq
In our stellar models, $L_{\rm wave}$ is calculated as described in \autoref{imp}, $T_{\rm shell}$ is calculated from equation \ref{t2} (with the $r$ locations corresponding to the edge of the burning shell, and $T_{\rm shell}=0$ for core burning phases), and $x_\nu$ is the integral in the exponent of equation \ref{fneu}. Our code calculates the transmission coefficients of all evanescent zones overlying the wave generation zone, and $T_{\rm min}$ is the minimum transmission coefficient found in each model. Note that in the limit of no damping in the core ($x_\nu = T_{\rm shell}=0$), all of the wave energy  escapes into the envelope.

\subsection{Wave Damping via Radiative Diffusion}
\label{wavedamp}

Away from evanescent regions, waves are well approximated by the WKB limit, in which the wave damping rate is 
\beq
\label{damp}
\frac{\dot{L}_{\rm wave}}{L_{\rm wave}} = \gamma = k_r^2 K \,
\eeq
where $K$ is the thermal diffusivity
\beq
\label{diff}
K = \frac{16 \sigma_{\rm SB} T^3}{3 \rho^2 c_p \kappa} \,
\eeq
and $\sigma_{\rm SB}$ is the Stephan-Boltzmann constant, $T$ is temperature, $c_p$ is specific heat at constant pressure, and $\kappa$ is the Rosseland mean opacity. We find radiative diffusion is only important in the envelope of the stars where waves are well approximated as WKB acoustic waves with $k_r = \omega/c_s$. In this limit, the waves travel at group speed $v_g = c_s$ and we can define a damping length $l_{\rm damp} = v_g/\gamma = c_s^3/(\omega^2 K)$. Then the waves damp after traversing a mass $M_{\rm damp} = 4 \pi \rho r^2 l_{\rm damp}$, which evaluates to 
\beq
\label{Mdamp}
M_{\rm damp} = \frac{4 \pi \rho r^2 c_s^3}{\omega^2 K} = \frac{3 \pi \rho^3 r^2 c_s^3 c_p \kappa}{4 \sigma_{\rm SB} \omega^2 T^3} \, .
\eeq
Equating $M_{\rm damp}$ with a the mass in one scale height roughly reproduces the damping criterion of equation 7 of \cite{quataert:12}.

As waves propagate upward, they damp out at a rate
\beq
\label{dldm}
\frac{ d L_{\rm wave}}{d M} = -\frac{L_{\rm wave}}{M_{\rm damp}} \, .
\eeq
In our numerical implementation, after calculating the fraction of energy escaping into the envelope as acoustic waves, we damp out wave energy such that the decrease in wave luminosity $L_{\rm wave}$ across a cell of mass $\Delta m$ is
\beq
\Delta L_{\rm wave} = -\frac{L_{\rm wave} \Delta m}{M_{\rm damp}} \, . 
\eeq
The corresponding amount of heat added to the cell per unit mass per unit time is thus 
\beq
\epsilon_{\rm heat} = \frac{L_{\rm wave}}{M_{\rm damp}} \, . 
\eeq

The most important feature of equation \ref{Mdamp} is its strong dependence on density (other factors tend to somewhat cancel each other out). As waves propagate out of the core and into the envelope, the density drops by several orders of magnitude just outside the helium core (see \autoref{fig:WaveDamp15Msun}). At this location, the damping mass drops from a value that is orders of magnitude larger than the interior mass to a value orders of magnitude smaller than the exterior mass. This means that waves are essentially undamped below this region, but totally damped when they propagate into this region. The waves tunneling out of the core will thus deposit all their energy as heat near the base of the hydrogen envelope.

A simplification of our method is to ignore the wave propagation time between excitation and damping. This approximation is reasonable because propagation time scales to the base of the hydrogen envelope are hours to days, whereas stellar evolution timescales are months to years for waves excited during Ne/O burning. However, the propagation delay will need to be included to model wave heating during late O-shell burning and Si burning, when wave propagation times are comparable to evolution time scales.

\section{Effects of Magnetic Fields}
\label{magnetic}

Magnetic fields larger than a critical value \citep{fuller:15,lecoanet:17}
\beq
B_c \sim \sqrt{\frac{\pi \rho}{2}} \frac{\omega^2 r}{N} \,
\eeq
will prevent gravity wave propagation in stably stratified regions, converting gravity waves into Alfv\'en-like waves, with a slight dependence on magnetic field geometry. In \autoref{fig:15MsunBc}, we plot the value of $B_c$ in our model during core O-burning for the wave frequency $\omega_{\rm wave} = 5 \times 10^{-3}$. At this stage, a magnetic field of $B \gtrsim 2 \! \times \! 10^{7} \, {\rm G}$ in the radiative C/O/Ne shell above the convective core would be sufficient to suppress gravity wave propagation and alter wave heating. This magnetic flux is comparable to that found in young pulsars, magnetic white dwarfs, and magnetic Ap/Bp stars, and may plausibly exist in massive stellar cores.

\begin{figure}
\begin{center}
\includegraphics[scale=0.36]{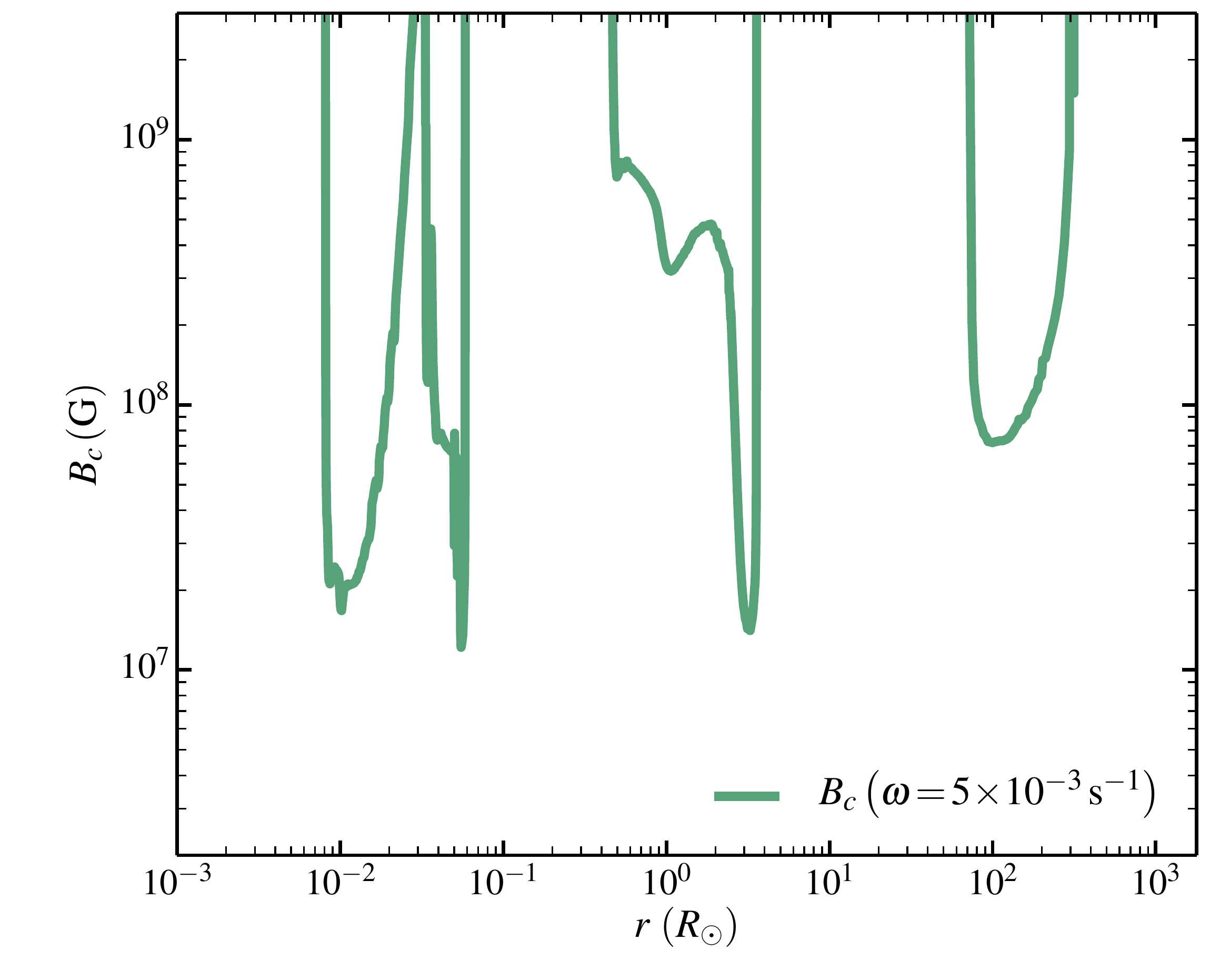}
\end{center} 
\caption{ \label{fig:15MsunBc} 
Minimum radial magnetic field strength $B_c$ needed to suppress convectively excited gravity waves of frequency $\omega_{\rm wave} = 5\times 10^{-3} \, {\rm rad}/{\rm s}$ during core oxygen burning. In the radiative core surrounding the oxygen burning shell, a field strength $B_c \! \sim \! 2 \! \times \! 10^7 \, {\rm G}$ is required to suppress waves, a magnetic field strength comparable to typical magnetic white dwarfs, and magnetic flux comparable to young pulsars. }
\end{figure}

Unfortunately, it is very difficult to estimate core magnetic field strengths of massive stars. If magnetic fields generated during previous convective core burning phases survive beyond C-burning, they can account for the required magnetic flux. There is evidence in lower mass stars that core fields frequently survive after being generated by a main sequence core dynamo (see discussion in \citealt{stello:16,cantiello:16}), although it is not clear whether they would survive subsequent convective phases like those in massive stars.

If strong core fields do exist, gravity wave energy will be converted in Alfv\'en wave energy within the core. The fate of this energy is uncertain and depends on the global magnetic field topology. However, we speculate field strengths will be much smaller at larger mass coordinates with lower densities. This may cause Alfv\'en waves to damp in the outer core before reaching the hydrogen envelope. In this case, wave heating energy will probably have a negligible affect on the stellar structure due to the large binding energy of the core relative to the wave energy, and a pre-SN outburst would be suppressed.

\end{document}